\documentclass{article}

\usepackage[english]{babel}

\usepackage[letterpaper,top=2cm,bottom=2cm,left=3cm,right=3cm,marginparwidth=1.75cm]{geometry}

\usepackage{amsmath}
\usepackage{graphicx}
\usepackage[colorlinks=true, allcolors=blue]{hyperref}
\usepackage[printonlyused]{acronym}
\usepackage{subcaption}
\usepackage{caption}
\usepackage{float}
\usepackage{geometry}
\usepackage{enumitem}
\usepackage{lipsum}
\usepackage{algorithm}
\usepackage{algpseudocode}
\usepackage{setspace}
\usepackage{booktabs}
\usepackage{threeparttable}
\usepackage{multirow}
\usepackage{authblk}
\date{} 

\title{Emerging Paradigms in the Energy Sector: Forecasting and System Control Optimisation}
\author[1]{Dariush Pourkeramati}
\author[1]{Gareth Wadge}
\author[1]{Rachel Hassall}
\author[1]{Charlotte Mitchell}
\author[1]{Anish Khadka}
\author[1]{Shiwang Jaiswal}
\author[2]{Andrew Duncan}
\author[2]{Rossella Arcucci}

\affil[1]{SSE plc, UK}
\affil[2]{Imperial College London, UK}

\begin{document}
\maketitle

\begin{abstract}

The energy sector is experiencing rapid transformation due to increasing renewable energy integration, decentralisation of power systems, and a heightened focus on efficiency and sustainability. With energy demand becoming increasingly dynamic and generation sources more variable, advanced forecasting and optimisation strategies are crucial for maintaining grid stability, cost-effectiveness, and environmental sustainability. This paper explores emerging paradigms in energy forecasting and management, emphasizing four critical domains: Energy Demand Forecasting integrated with Weather Data, Building Energy Optimisation, Heat Network Optimisation, and Energy Management System (EMS) Optimisation within a System of Systems (SoS) framework.

Leveraging machine learning techniques and Model Predictive Control (MPC), the study demonstrates substantial enhancements in energy efficiency across scales — from individual buildings to complex interconnected energy networks. Weather-informed demand forecasting significantly improves grid resilience and resource allocation strategies. Smart building optimisation integrates predictive analytics to substantially reduce energy consumption without compromising occupant comfort. Optimising CHP-based heat networks achieves cost and carbon savings while adhering to operational and asset constraints. At the systems level, sophisticated EMS optimisation ensures coordinated control of distributed resources, storage solutions, and demand-side flexibility.

Through real-world case studies we highlight the potential of AI-driven automation and integrated control solutions in facilitating a resilient, efficient, and sustainable energy future.
\end{abstract}

\section{Introduction}

The energy sector is undergoing a transformative shift driven by the increasing integration of renewable energy, the decentralisation of power systems, and the imperative for improved efficiency and sustainability. As energy demand fluctuates and generation becomes more variable due to renewables, advanced forecasting and system control optimisation techniques have emerged as critical tools for ensuring grid stability, cost-effectiveness, and environmental sustainability \cite{turk2017digitalization}. 
 These emerging paradigms are particularly relevant in key areas such as building energy optimisation, heat network optimisation, energy demand forecasting with integrated weather data, and \ac{EMS} optimisation in a systems-of-systems context.

Accurate energy demand forecasting is a fundamental requirement for modern power systems, enabling efficient resource allocation, grid stability, and cost-effective energy management. Forecasting energy consumption at various levels — ranging from individual meters to aggregated portfolio scales — is essential for utilities, energy providers, and businesses seeking to optimise operations and reduce inefficiencies~\cite{sinn2014energy}. However, energy demand is inherently complex and influenced by multiple factors, including weather conditions, sector-specific consumption patterns, and operational variations~\cite{kaur2020energy}. Traditional forecasting methods often struggle to balance accuracy and scalability across different organisational levels~\cite{mystakidis2024energy}.

Building energy optimisation is central to reducing energy consumption and improving efficiency in urban environments. Buildings account for approximately 34\% of global energy consumption and contribute to 37\% of global carbon dioxide emissions \cite{unep22}. A significant portion of this energy is used for \ac{HVAC} systems, making them a critical target for optimisation in the fight against climate change. As energy prices rise, stricter carbon regulations come into effect, and building occupancy patterns become increasingly variable, building operators face increasing pressure to reduce costs while maintaining occupant comfort and adhering to sustainability goals.
Advanced control strategies and AI-driven predictive analytics are now being used to dynamically manage heating, cooling, lighting, and occupancy-based energy usage.

Similarly at the district or campus scale, effective heat network optimisation ensures that existing fossil fuel based heat networks and the low carbon networks of the future can be operated in a way that maximises cost and carbon efficiency ~\cite{sun2022bcs}. Both areas can benefit from real-time forecasting and optimisation techniques that ensure the most efficient energy use while maintaining occupant comfort and operational reliability~\cite{yu2023flexible}.

In a broader sense, \ac{EMS} optimisation in a \ac{SoS} context plays a fundamental role in balancing supply and demand across complex, multi-layered infrastructures. Modern \ac{EMS} approaches integrate distributed energy resources, grid-scale storage, demand-side management, and real-time market signals to optimise system-wide performance. Leveraging \ac{MPC} and machine learning, these systems can dynamically adjust energy flow, improve resilience, and support a seamless transition to sustainable energy ecosystems~\cite{meng2025enhanced}.

This paper explores these emerging paradigms in energy forecasting and system control optimisation, analysing how advancements in AI, automation, and integrated modelling are shaping the future of energy management. By examining cutting-edge methodologies and their applications we demonstrate how a common digital-twin plus optimiser workflow can scale from a single building, to a district heat-network, and finally to a multi-vector micro-grid, thus contributing to a more efficient, resilient, and sustainable energy landscape.
\\\\
To this end, we focus on four pillars of energy system management and optimisation:

\begin{itemize}
\item Energy Demand Forecasting with Integrated Weather Data: accurate energy forecasting is the foundation for all optimisation efforts. Weather plays a critical role in energy consumption patterns, especially for heating, cooling, and renewable energy integration. A well-optimised system starts with precise forecasting to anticipate demand fluctuations. 

\item Building Energy Optimisation: once demand forecasting is established, the next step is optimising energy consumption at the building level whilst maintaining comfort. This involves strategies like \ac{HVAC} control, smart sensors, and occupancy-based energy management, reducing unnecessary energy use and improving efficiency at the micro-level;

\item Heat Network Optimisation: \ac{CHP} systems ensure efficient generation and distribution of heat and power. \ac{CHP} optimisation plays a key role in district heating, industrial applications, and decentralised energy production, bridging the gap between local energy demand and supply;

\item Energy Management System Optimisation in a \ac{SoS} Context:  system-wide coordination is necessary to integrate individual buildings, \ac{CHP} systems, and renewable energy assets into a unified, intelligent grid. Optimising energy at the macro-level ensures resilient, flexible, and sustainable energy distribution across interconnected systems and a \emph{whole-systems} approach. By leveraging decentralised control, data integration, and adaptive strategies, it ensures optimal resource allocation, resilience, and sustainability.
\end{itemize}

The remaining body of this paper is structured as follows:
Section~\ref{section:relatedwork} reviews related work in Energy Demand Forecasting with Integrated Weather Data, Building Energy Optimisation, \ac{CHP} Heat Network Optimisation, and Energy Management System Optimisation in a \ac{SoS} Context, highlighting recent contributions and research gaps. Section~\ref{section:chrono}, Section~\ref{section:BeOpt},  Section~\ref{section:CHOP} and Section~\ref{section:ems} discuss each of the four pillars in detail. 
Each section presents methodology and implementation of the approach and provides an analysis and discussion of the models for real test cases.
The paper concludes with Section~\ref{section:conclusion}.

\section{System Overview and Existing Approaches}\label{section:relatedwork}

The increasing complexity of modern energy systems, driven by the transition to renewable energy sources, grid decentralisation, and digitalisation, necessitates advanced forecasting techniques and optimisation strategies. Traditional methods such as time series forecasting, rule-based energy management, and heuristic optimisation have been widely used but are being increasingly supplemented by AI-driven techniques such as deep learning, reinforcement learning, and digital twin technology.

This section presents some related works implementing both standard and emerging methodologies in Energy Demand Forecasting with Integrated Weather Data, Building Energy Optimisation, \ac{CHP} Optimisation, and Energy Management System Optimisation in a \ac{SoS} Context, highlighting recent contributions and research gaps.

\subsection{Energy Demand Forecasting with Integrated Weather Data}

Electric load forecasting plays an important role in the operation, planning, and activities of the energy market \cite{markovic2023machine}. Grid operators and utilities rely on accurate demand forecasts to balance generation with consumption, plan capacity, and ensure reliable service~\cite{hong2016probabilistic,willis2002spatial}. Even modest improvements in forecast accuracy can yield significant benefits: operating the system closer to optimal improves efficiency, profitability, and grid stability~\cite{hong2014global}.  Forecasting in building energy systems encompasses a wide variety of temporal horizons and spatial granularities, ranging from minute-level \ac{HVAC} predictions in individual rooms, to day ahead forecasts for planning battery charge/discharge scheduling.   Traditionally utilities have employed basic statistical models for time-series, such as \ac{ARIMA} and Holt-Winters exponential smoothing \cite{hyndman2009density,taylor2003short}.  These been effective models for short-term forecasting in energy scenarios, due to their due to their simplicity, theoretical guarantees and low computational requirements.   However, these models assume varying degrees of linearity and stationarity in the data, which limits their ability to adapt to changing consumption patterns or non-linear interactions, particularly those induced by weather or behavioural trends.
\\\\
Recent research has explored machine learning models that do not require strong assumptions on data distribution. Tree-based ensemble methods (e.g., Random Forests, \ac{XGBoost}, \ac{LightGBM}) and deep learning architectures (e.g., \ac{LSTM}, \ac{RNN}s and Transformers), \cite{taieb2014gradient,liu2014hybrid,marino2016building, zheng2014time} have demonstrated significant effectiveness in scenarios with high-frequency data and non-trivial feature dependencies. These models are particularly effective at capturing complex nonlinear relationships and intricate weather–load interactions, particularly modelling temperature, humidity, solar radiation, and calendar effects.
\\\\
However, model selection remains highly problem-specific. Empirical comparisons over various studies confirm that no single algorithm dominates across all tasks and horizons - echoing the “no free lunch” theorem~\cite{wolpert1997no}. This motivates the  use of model ensembles, which aggregate the strengths of diverse learners. Simple ensembles, such as unweighted averaging or voting, often yield stable performance improvements. More sophisticated methods, such as stacked generalization, residual modeling, and Bayesian model averaging, aim to dynamically weight component models based on their context-specific strengths. These techniques are particularly advantageous in energy forecasting, where different regimes (e.g., weekdays vs weekends, seasons, holidays) may favour different modelling strategies~\cite{makridakis2018m4, hong2016probabilistic}.
\\\\
In practical deployments, the choice of model is often shaped by constraints such as interpretability, data availability, and computational efficiency. While black-box models may deliver strong predictive performance, they are frequently viewed with apprehension in operational settings, where transparent decision rules and predictable resource demands are essential \cite{rudin2019stop}. Hybrid approaches that incorporate physical constraints or embed physics-informed structures into machine learning models are gaining traction, offering a balance between accuracy and transparency \cite{guo2024grey,von2021informed}.

\subsubsection{Bottom-up versus Top-Down Energy Forecasting}

Although there is a substantial body of research on electricity demand forecasting, the majority of existing studies predominantly focus on predicting total electricity consumption using time series models that rely primarily on previous aggregate demand values. Energy demand is often organized hierarchically by region, sector, or customer contracts. The widespread deployment of smart metering infrastructure and the increasing availability of high-resolution electricity usage data have opened new avenues to improve forecast accuracy by leveraging detailed consumption patterns at increasingly finer granularities.
\\\\
In response to these technological advances, a growing number of studies have adopted what is known as a ``bottom-up'' forecasting strategy. This method involves building individual forecasting models for disaggregated units - such as households or small business premises - and subsequently aggregating their predictions to estimate total electricity demand. While a bottom-up approach enables the incorporation of granular consumption data and offers the potential for more personalised and localised predictions, it also introduces several challenges. These include increased model complexity, the risk of error propagation through the aggregation process, and the potential for inefficiencies when modelling a large number of sub-units independently. As a result, the bottom-up methodology, despite its promise, may not always lead to optimal forecasting performance at the system level \cite{anand2023bottom}.
\\\\
Energy demand is often organised hierarchically by region, sector, or customer contracts. Forecasts at each level should ideally be coherent (i.e., sub-level forecasts sum to the top-level). Traditional strategies include bottom-up (sum component forecasts) and top-down (disaggregate an aggregate forecast). Research shows both can be beneficial depending on correlation among lower-level series and data availability~\cite{hyndman2011optimal}. Advanced reconciliation methods (e.g., MinT) also achieve consistent forecasts across the hierarchy~\cite{hyndman2018forecasting}. Our work addresses multi-scale forecasting, though we focus primarily on bottom-up with an optional top-down comparison.

\subsection{Energy system optimisation under uncertainty}
Energy systems operate in environments characterised by multiple sources of uncertainty: fluctuating weather conditions, variable occupancy, intermittent on-site generation (e.g., solar PV), and dynamic pricing or carbon intensity signals. Consequently, decision-making processes for control and scheduling must be robust to forecast errors and adaptable to new information.
\\\\
\ac{MPC} has emerged as a dominant paradigm for such contexts. \ac{MPC} operates by solving a finite-horizon optimisation problem at each time step, using a forecast of future fluctuations (e.g., demand, weather, prices) to plan statistically optimal control actions. Only the first control action is implemented, and the process is repeated as new data becomes available. This receding-horizon strategy allows \ac{MPC} to respond adaptively to forecast revisions and evolving system dynamics. \ac{MPC} frameworks are particularly well-suited for \ac{HVAC} control, thermal storage scheduling, and \ac{CHP} dispatch, where system constraints and time-coupled dynamics are prominent~\cite{rawlings2017model}.
\\\\
Beyond \ac{MPC}, several planning strategies incorporate uncertainty directly into the optimisation problem. Stochastic programming approaches formulate multi-scenario optimisation problems, where decision variables are optimised to perform well on average or worst-case scenarios, across a distribution of possible futures. Robust optimisation techniques adopt a more conservative strategy, seeking solutions that perform adequately under the worst-case realization within a predefined uncertainty set~\cite{ben2009robust}. These approaches are particularly relevant in mission-critical systems where over-optimistic assumptions could result in infeasibility or comfort violations.
\\\\
More recently, \ac{RL} has gained attention for real-time energy management and control,  particularly in multi-agent or hierarchical settings.   \ac{RL} algorithms can learn optimal policies through interaction with the environment, without requiring explicit system models. However, \ac{RL} methods often require extensive exploration, carry stability concerns during training, and may struggle with hard constraints, making their integration into real-world systems non-trivial~\cite{wei2017deep}. Hybrid approaches that combine \ac{RL} with \ac{MPC} (e.g., \ac{RL}-based policy initialization or switching strategies) are an area of active research.
\\\\
Across all these methods, the integration of forecast uncertainty quantification, via probabilistic forecasts remains essential for effective planning. Deterministic forecasts may suffice in stable regimes, but under uncertainty, planning tools must reason over possible futures to ensure reliability and cost-effectiveness. This interplay between forecasting and control is a key focus of recent research in energy-aware AI systems.

\subsubsection{Building Energy Optimisation}

Building energy systems present a rich testbed for optimisation under uncertainty, where fluctuating occupancy, weather, and electricity prices influence \ac{HVAC} operation, lighting, and storage decisions. Traditionally, energy consumption in buildings has been controlled using rule-based strategies, where predefined operational rules adjust heating, cooling, and lighting based on occupancy and time schedules \cite{clarke2007energy}.
\\\\
To move beyond static heuristics, \ac{LP} and \ac{MILP} have been widely applied for demand-response optimisation and \ac{HVAC} control. These approaches offer improved energy efficiency but often rely on deterministic assumptions \cite{fumo2014review}.  
\\\\
More recently, \ac{DL} models such as \ac{LSTM} networks have been employed to capture temporal demand patterns and improve predictive accuracy in smart buildings \cite{runge2021review}.  In \cite{yang2023auto}, a Modified \ac{KF} was integrated with \ac{LSTM} networks for real-time energy prediction, significantly improving accuracy. Building on these predictive capabilities, \cite{arsecularatne2024digital} demonstrates how digital twin frameworks, combining IoT sensor data with AI models, enable real-time, data-driven control and optimisation of building energy systems.

Several reviews and studies have highlighted the growing role of \ac{ML} techniques in improving energy efficiency, forecasting, and control. For instance, \cite{Das2024} provide a comprehensive overview of \ac{ML} applications for energy-efficient buildings, while \cite{Villano2024} focus on \ac{ML} and deep learning methods for building simulation and energy management. \cite{Bahrami2023} extend this discussion to power electronics, emphasising \ac{ML}'s potential in control and optimisation tasks. \cite{Markovic2023} discuss ML integration into power distribution networks, outlining both challenges and opportunities. In terms of control frameworks, \cite{Dagdougui2020} and \cite{Arroyo2022} compare predictive and optimal control strategies for building energy systems. Complementing these, \cite{Dhaigude2024} showcase the importance of real-time analytics in \ac{BMS}, while \cite{Tian2024} explore the fusion of \ac{LSTM} networks with \ac{KF}s, relevant for sensor-based applications. Finally, \cite{Dmitrewski2022} presents CntrlDA, a real-time control system that adapts \ac{HVAC} operations to achieve optimal indoor temperature regulation.


\subsubsection{Combined Heat and Power (\ac{CHP}) Optimisation}

Much like buildings, \ac{CHP} systems also operate under uncertainty but face additional thermodynamic and interdependency constraints. Traditional \ac{CHP} scheduling is formulated using thermodynamic models and \ac{MILP} to determine cost-optimal dispatch plans that balance electricity and heat demands \cite{lund2010role}.  Unit commitment models are often used to coordinate on/off states and load levels, ensuring reliability while minimising operational costs \cite{kaur2023economic}.
\\\\
\ac{MPC} has been increasingly adopted in microgrid contexts to manage \ac{DER}, including \ac{CHP} units.  For example, \cite{joshal2023microgrids} shows how \ac{MPC} enhances power quality, stability, and energy management through real-time receding-horizon optimisation. Recently, strategies combining distributed and data-driven control have been explored. In \cite{zhang2023multi}, the authors propose a multi-agent deep reinforcement learning framework for privacy-preserving scheduling across interconnected energy hubs with renewable generation. This highlights the frontier where \ac{CHP} optimisation and management intersects with decentralised control, real-time learning, building on the techniques described earlier in this section.

\subsubsection{Energy Management System Optimisation in a \ac{SoS} Context}


Buildings and campuses are increasingly adopting on-site generation such as solar or wind, alongside \ac{BESS}, to reduce both operational costs and carbon dioxide emissions \cite{unep20202022}. However, effectively orchestrating these assets can be challenging given time-varying energy tariffs, carbon intensity signals, and local constraints on peak demand or interconnectivity limits. An \ac{EMS} plays a crucial role in coordinating generation and storage to meet load requirements at minimal cost while adhering to stakeholder goals around sustainability and comfort.
\\\\
Recent advances in \ac{EMS} design emphasise the synergy of: physically grounded models (digital twins) that capture real-world thermodynamic and electrical behaviours; optimisation-based controllers (e.g. \ac{MILP}) that can systematically explore cost and carbon trade-offs, scheduling energy flows over a user-defined horizon; scenario analyses to test how the EMS performs under diverse tariff structures and to highlight the business case for flexible operation \cite{allwyn2023comprehensive}. Moreover, as the concept of \ac{SoS} gains traction in the energy domain, it becomes vital to ensure that local EMS strategies also integrate with broader grid objectives and multi-site coordination \cite{maier1998architecting}. Likewise, microgrid deployments often require the same optimisation logic to be extended to include islanding and advanced grid-support functionalities \cite{momoh2012smart, he2012enhanced}.

\section{Energy Demand Forecasting with Integrated Weather Data: From Individual Meters to Portfolio Level}\label{section:chrono}

Accurate energy demand forecasting is critical for modern power systems, from granular consumption at individual meters up to aggregated portfolio levels.  Here, we introduce a novel weighted orchestrator ensemble framework for multi-scale energy demand forecasting that integrates weather data and automated \ac{MLOps}. The approach combines heterogeneous models (including \ac{XGBoost}, \ac{LightGBM}, and \ac{EMA}) in an ensemble, where a learned weighting ``orchestrator'' dynamically blends predictions to leverage each model's strengths. The system is designed to forecast at multiple organisational scales - Contract, Sector, District, and Portfolio - providing flexibility from individual \acp{MPAN} to an entire portfolio. An end-to-end pipeline is implemented to automate data ingestion, \ac{ETL} processing, \ac{MPAN} filtering, anomaly detection, model training, and deployment using \ac{MLOps} tools (Databricks, MLflow, Azure Data Factory, Unity Catalog). We demonstrate the methodology on two real-world case studies: (1) bottom-up forecasting for 300 aggregated \ac{MPAN}s and (2) sub-meter level forecasting within a facility. Results show improved week-ahead and year-ahead forecast accuracy compared to traditional single-model approaches, with the ensemble reducing errors (measured by \ac{MAE}, \ac{MAPE}, \ac{RMSE}) across all scales. We also discuss key advantages of this architecture (such as improved accuracy through ensemble synergy and scalable \ac{MLOps} integration) and potential pitfalls (including data drift, model-overfitting risks, and integration complexity). Finally, future directions are outlined, including incorporating advanced deep learning models (e.g., Temporal Fusion Transformers, N-HiTS) and probabilistic forecasting (e.g., TimeGPT foundation model), to further enhance the robustness and interpretability of energy demand forecasts.

Motivated by the above challenges and opportunities, this work introduces a comprehensive forecasting framework with the following key attributes:

\begin{itemize}
    \item \textbf{Weighted Ensemble Orchestrator:} We develop a novel ensemble strategy that dynamically orchestrates a set of forecasting models using optimised weights. The orchestrator automatically adjusts the contribution of each model (e.g., \ac{XGBoost}, \ac{LightGBM}, \ac{EMA}) based on recent performance, effectively creating a self-tuning hybrid model.

    \item \textbf{Multi-Scale Forecasting from Meters to Portfolio:} The proposed approach is inherently multi-scale, capable of forecasting at various hierarchical levels - from individual meter (\ac{MPAN}) level, to aggregated contract or sector level, up to an entire portfolio.

    \item \textbf{Integration of Weather and Anomaly Detection:} The methodology tightly integrates exogenous weather data (temperature, humidity, wind, etc.) into the forecasting models to improve accuracy. We also implement an automated data pipeline with anomaly detection and filtering to preprocess meter data.

    \item \textbf{Automated \ac{MLOps} Pipeline:} We demonstrate an end-to-end machine learning operations pipeline for continuous training, deployment, and monitoring of the forecasting models. Using tools such as Azure Databricks, MLflow, Azure Data Factory, and Unity Catalog, we show how the solution can be deployed in a real production environment.

    \item \textbf{Real-World Validation:} The approach is validated on two real-world case studies: (a) a bottom-up forecast of 300 \ac{MPAN}s where individual meter forecasts are aggregated, and (b) a sub-meter level forecast within a large facility. Results highlight the versatility of the framework, showing improvements in \ac{MAE}, \ac{MAPE}, and \ac{RMSE} compared to single-model baselines.
\end{itemize}

\subsection{Methodology and Implementation}\label{sec:method}
In this subsection, we describe the proposed forecasting methodology in detail. We first outline the overall data pipeline and system architecture, then explain the weighted ensemble orchestrator framework and the models involved. Next, we define the various forecasting scenarios across different aggregation levels (Contract, Sector, District, Portfolio) and their implications. Finally, we discuss the \ac{MLOps} components that support the model lifecycle and deployment.\\

The end-to-end data pipeline (from data ingestion to forecast output) involves:
\begin{itemize}
    \item \textit{Data Ingestion \& \ac{ETL}:} Automated extraction of meter consumption data and weather data, cleaning timestamps, handling missing values, and merging features.
    \item \textit{\ac{MPAN} Filtering:} Selection of active meters with sufficient history.
    \item \textit{Anomaly Detection:} Automatic removal or imputation of outliers and sensor errors, using statistical thresholds or isolation-forest-based approaches.
    \item \textit{Feature Store:} Storing processed features (lags, weather, calendar flags) for consistency in training and inference.
    \item \textit{Model Training:} Automated retraining of models (e.g., \ac{XGBoost}, \ac{LightGBM}) on each time series or aggregated level, scheduled via orchestration tools (Azure Data Factory or Databricks Workflows).
    \item \textit{Forecast Generation:} Running inference for week-ahead or year-ahead horizons, optionally aggregating bottom-up to portfolio or disaggregating top-down.
    \item \textit{Monitoring:} Logging forecasts vs. actuals in MLflow, triggering alerts if accuracy degrades.
\end{itemize}
This pipeline is designed for scalability (multiple series, parallel training) and traceability (each model run is tracked). 
We handle multiple organizational scales:
at a Contract-Level, we aggregate loads of buildings or assets under a specific contract. At a Sector-Level we group by customer industry type (schools, offices, etc.). At a District-Level we group by geographic location. and at Portfolio-Level we consider the total load across all meters.
A bottom-up approach provides a transparent way to sum from meters to higher levels; we also compare direct modelling of aggregate (top-down). Our pipeline can schedule either approach depending on user needs and data availability.

At the core of our methodology is a \textit{weighted ensemble orchestrator} that combines forecasts from multiple base models. Denoting each model’s forecast at time $t+h$ as $\hat{y}^{(m)}_{t+h}$ for $m=1,\dots,M$, the final ensemble forecast $\hat{y}^{(ens)}_{t+h}$ is:
\[
\hat{y}^{(\text{ens})}_{t+h} = \sum_{m=1}^{M} w_m\, \hat{y}^{(m)}_{t+h},
\]
where ${w_m}$ are non-negative weights summing to 1. We consider three main models: 
\textit{(1) \ac{XGBoost}}, 
\textit{(2) \ac{LightGBM}}, 
and \textit{(3) \ac{EMA}}. 
Weights are updated dynamically based on recent performance (e.g., inverse of rolling error), allowing the ensemble adapt as different models become more accurate under evolving conditions.

To ensure a production-grade environment, we incorporate Azure Databricks 
for scalable data preparation and distributed model training, MLflow 
for experiment tracking and model registry, ensuring reproducibility and version control. Azure Data Factory 
is used here for orchestrating \ac{ETL} and model training workflows on a defined schedule and Unity Catalog 
for data governance and lineage tracking, ensuring consistent feature definitions across training/inference.
This \ac{MLOps} setup automates retraining, continuous monitoring, and rapid deployment of new model versions when performance drifts, reducing the overhead of manual maintenance.\\

\subsection{Tests, Analysis and Discussion}\label{sec:testing}

We applied the above methodology in a real-world energy forecasting project to validate its effectiveness. Below we highlight the implementation details, focusing on data flows, model orchestration, and practical lessons learned.

As a test case, we forecasted electricity consumption for a portfolio of around 300 individual meters \acp{MPAN} spanning multiple sites and leveraging a combination of multiple machine learning models, naïve forecasting techniques, and an ensemble model to enhance predictive accuracy. The forecasts are trained and validated using half-hourly meter data collected throughout 2022, allowing for granular insight into usage patterns and temporal demand fluctuations. Weather data included temperature, humidity, wind speed, and solar irradiance. Each \ac{MPAN} was filtered for sufficient history and cleaned via anomaly detection.

\begin{figure}[h!]
    \centering
    \includegraphics[width=0.7\linewidth]{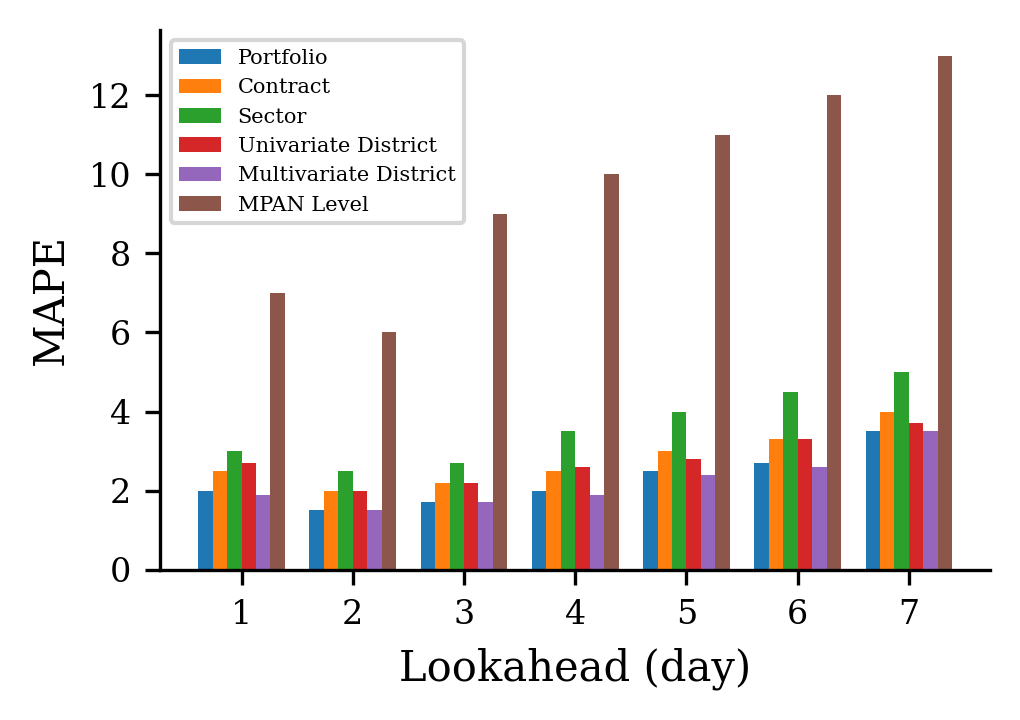}
    \caption{MAPE Comparison}
    \label{figure_manana1}
\end{figure}

We began by testing unsupervised learning techniques for clustering electricity demand profiles, but ultimately decided to use more explainable and interpretable groupings. To support our analysis, we developed three comparison scenarios based on aggregation by Contract, Sector, and District as shown in Figure~\ref{figure_manana1}. Each of these scenarios was designed to serve different use cases, offering distinct advantages depending on the context. Additionally, we incorporated weather data exclusively into the District-level forecast, as this level was most sensitive to geographic and climatic variations. We trained an \ac{XGBoost}, \ac{LightGBM}, and \ac{EMA} model per \ac{MPAN}, then aggregated forecasts to sector, district, and finally to the total portfolio. For week-ahead forecasts, we updated weights daily using rolling errors, allowing the ensemble to shift emphasis if weather changed or new usage patterns emerged.

\begin{figure}[H]
    \centering
    \begin{subfigure}[b]{0.5\textwidth}
        \includegraphics[width=\textwidth]{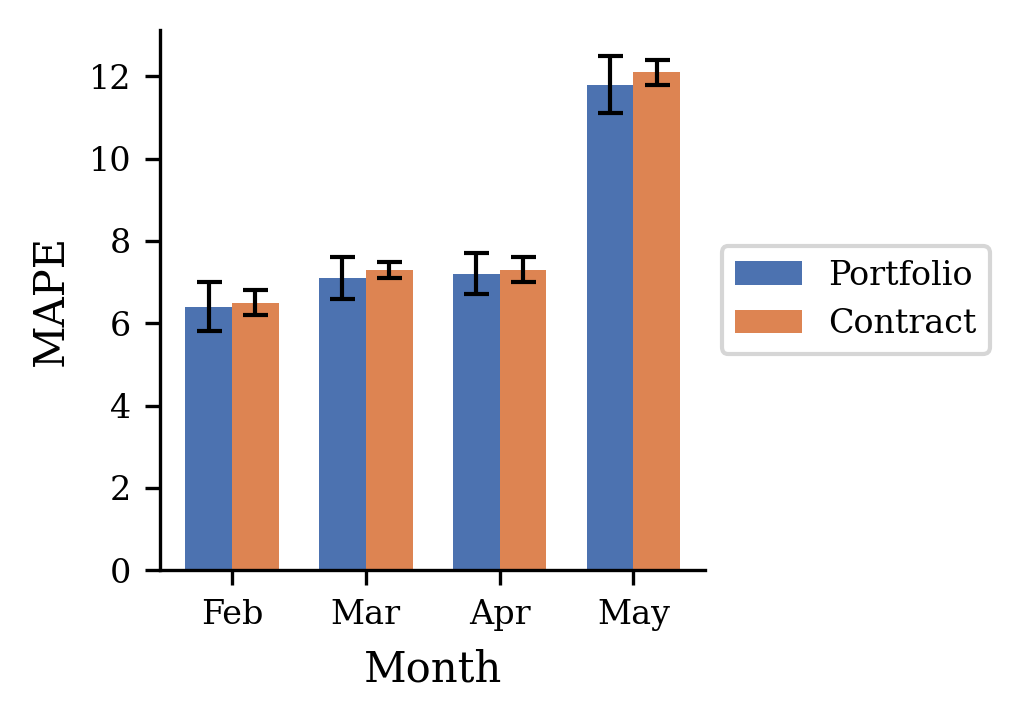}
        \caption{Contract level}
        \label{fig:a}
    \end{subfigure}
    \hfill
    \begin{subfigure}[b]{0.5\textwidth}
        \includegraphics[width=\textwidth]{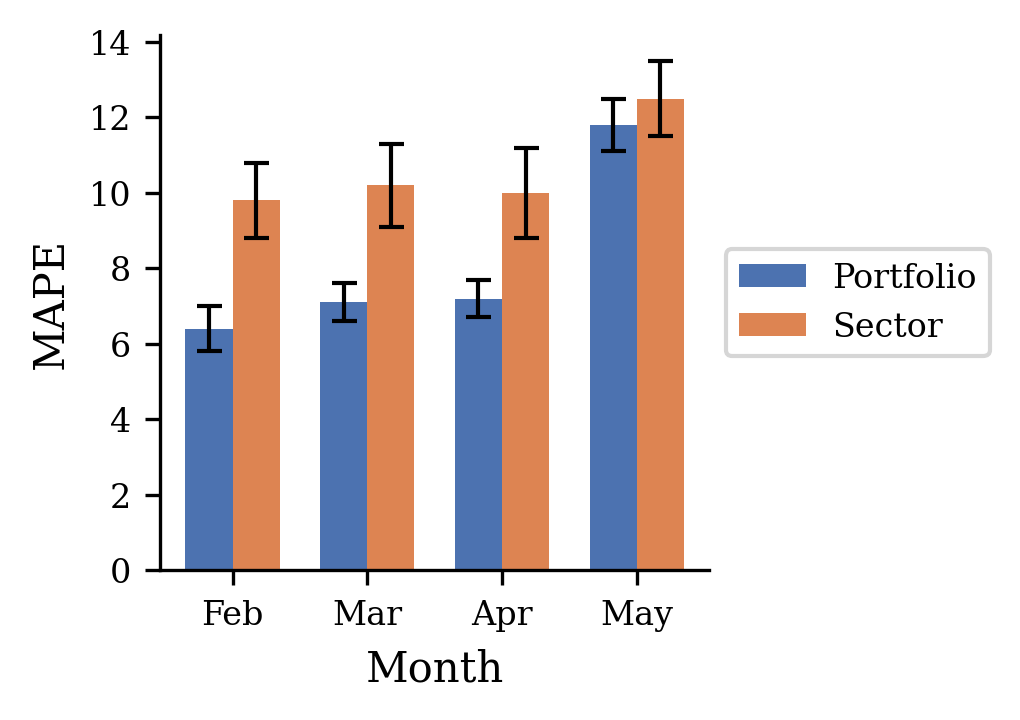}
        \caption{Sector level}
        \label{fig:b}
    \end{subfigure}
    \hfill
    \begin{subfigure}[b]{0.5\textwidth}
        \includegraphics[width=\textwidth]{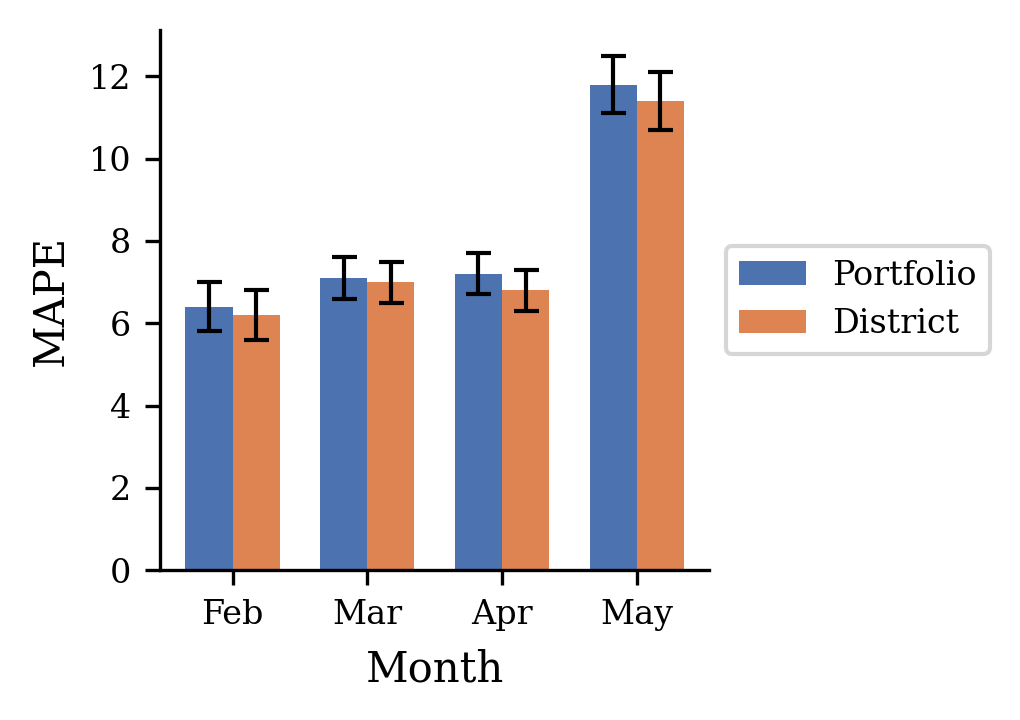}
        \caption{District level}
        \label{fig:c}
    \end{subfigure}
    \caption{\ac{MAPE} at (a) Contract level, (b) Sector level , and (c) District level.}
    \label{fig_MAPE_manana}
\end{figure}

As shown in Figure~\ref{fig_MAPE_manana} (a),
at the contract level, our findings showed no significant improvement in forecasting accuracy compared to other aggregation methods. However, this approach demonstrated low uncertainty and low model complexity, making it relatively simple to implement and manage. The contract-level aggregation also offered a moderate lifecycle in terms of its relevance and applicability over time. While the accuracy gains were limited, the potential value lies in enabling better probabilistic forecasting. This level of aggregation carries a small risk of error accumulation, but its simplicity and stability make it a useful option for scenarios where transparency and manageability are prioritised. 

At the sector level, our findings in Figure~\ref{fig_MAPE_manana} (b)
indicated relatively low forecasting accuracy accompanied by high uncertainty and increased model complexity. Despite these limitations, the sector-based approach demonstrated a long lifecycle, requiring minimal retraining over time, which makes it suitable for long-term monitoring and analysis. The key value of this approach lies in its potential for anomaly detection and the generation of inferred values in cases where direct data may be missing or incomplete. Moreover, sector-level aggregation appears to be less prone to producing unrealistic or ``hallucinated'' results, especially when data quality is poor, making it a robust option for handling noisy or sparse datasets.

Also, at the district level, our findings in Figure~\ref{fig_MAPE_manana} (c) shows slightly better forecasting accuracy compared to other aggregation methods, along with moderate uncertainty and complexity. This approach also benefits from a long lifecycle, requiring relatively infrequent retraining. One of the key strengths of district-level aggregation is its potential to serve as a viable candidate for bottom-up forecasting, particularly when combined with additional contextual data such as weather. While the current performance is promising, the district-level model also presents more room for improvement, making it a strong foundation for future enhancements and more sophisticated forecasting strategies.

\begin{figure}[h!]
    \centering
    \includegraphics[width=1\linewidth]{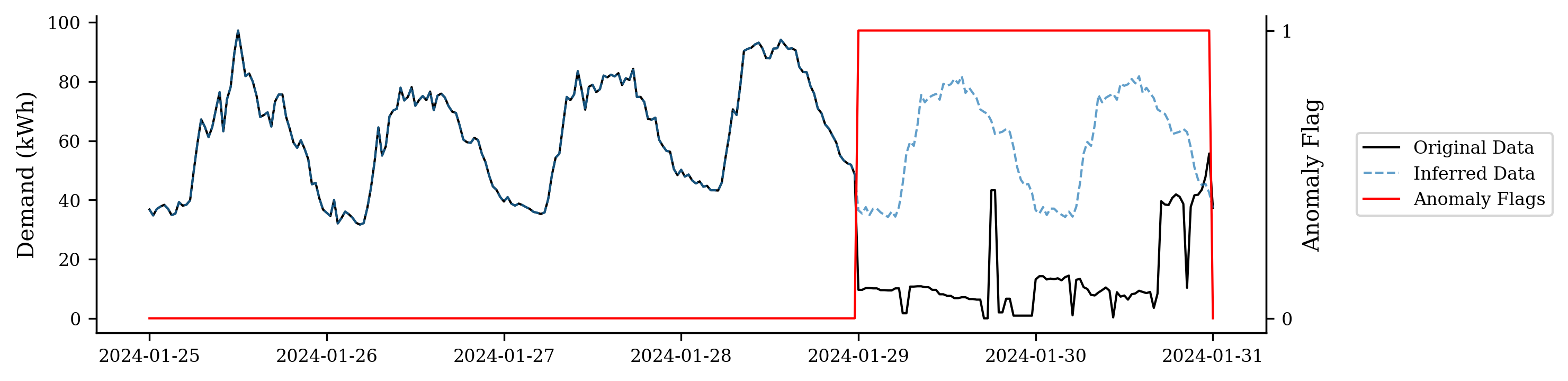}
    \caption{Model performance}
    \label{figure_manana2}
\end{figure}

Improving the performance of all models was a critical step, particularly given the challenges posed by poor data quality. These improvements were essential to ensure the effective use of machine learning models, which are highly sensitive to data inconsistencies and noise. The enhancements not only stabilised model outputs but also made the forecasting framework more robust and reliable. From a practical standpoint, this improved performance proved valuable for supporting estimation processes, especially in scenarios with incomplete or uncertain data as in Figure~\ref{figure_manana2}. While notable progress has been made, there remains significant room for further refinement, suggesting ongoing potential for optimisation and increased forecasting accuracy.

\section{Building Energy Optimisation}\label{section:BeOpt}
Optimising \ac{HVAC} operations through building energy optimisation offers both financial and environmental benefits that make it a compelling solution for addressing modern energy challenges. By intelligently aligning \ac{HVAC} operations with low-cost energy periods and reducing unnecessary heating or cooling, optimisation has demonstrated the potential to reduce energy costs by up to 25\%, resulting in substantial financial savings for building operators. Additionally, the system’s ability to dynamically respond to periods of low carbon intensity in the energy grid enhances energy efficiency, significantly reducing the carbon footprint of buildings. This aligns with global climate goals and supports compliance with increasingly stringent environmental regulations.

Beyond cost and carbon reductions, building energy optimisation ensures that internal temperatures remain consistently within the desired range, enhancing occupant comfort and productivity by avoiding overcooling or overheating. Automating \ac{HVAC} operations further reduces manual interventions, minimizes energy wastage, and extends the lifespan of \ac{HVAC} equipment by preventing overuse or reactive operational based on short-term local conditions. Once implemented, the approach can be scaled across multiple buildings, making it an effective tool for large-scale energy management in smart cities or building portfolios. This combination of financial savings, environmental impact, and operational efficiency underscores the transformative potential of building energy optimisation in modern energy optimisation.

While techniques for optimising \ac{HVAC} and building energy have previously been proposed, the real-world deployment of optimisation at the building level is challenging due to the dynamic interplay of several factors:

\begin{itemize}
    \item \textbf{Heterogeneity in building construction and operation:} Buildings have unique thermal characteristics influenced by structural properties such as insulation, thermal mass, and the size and placement of windows. These factors determine how the building responds to external weather conditions and internal stimuli including occupancy and \ac{HVAC} operations.
    \item \textbf{Variable external disturbances:} External conditions, including weather metrics like external temperature and solar irradiance, vary significantly across seasons, days, and even hours, requiring real-time adjustments to \ac{HVAC} operations.
    \item \textbf{Conflicting objectives:} The optimisation must balance cost efficiency, carbon emission reduction, and occupant comfort, which often conflict. For example, reducing energy usage might result in temperature deviations that affect comfort.
    \item \textbf{Energy market complexity:} Energy prices and carbon intensity fluctuate dynamically, introducing further variability into operational decisions.
    \item \textbf{Initial deployment window:} The statistical and machine learning models that drive a building energy optimisation system require a minimum amount of data before they can generate sufficiently accurate predictions to support effective control. Reducing this ``ramp-up'' phase  is essential for the scalable deployment of building optimisation systems across multiple sites.
\end{itemize}

The proposed approach aims to mitigate these limitations through the use of adaptive grey-box models, multi-objective optimisation with hyper-parameter search for objective weights, and high accuracy weather forecasts. The entire model and optimisation process utilise production-grade big-data tools and \ac{MLOps} paradigms, ensuring scalability and reducing bespoke modelling for different buildings.

\subsection{Methodology and Implementation}


A fundamental aspect of achieving an effective optimisation at the building-level is the selection of the thermal modelling approach, which encapsulates how the building responds to inputs from the \ac{HVAC} and external disturbances.
For building thermal modelling, we evaluated three approaches: white box, black box, and grey-box models ~\cite{buildingmpc}, each offering varying degrees of transparency and practicality. The white box model, while highly accurate and interpretable, requires exhaustive knowledge of building structures and geometry,  and is computationally expensive, making it infeasible for real-time applications in complex systems and cost-prohibitive for retrofitting to existing buildings. Black box models, such as neural networks, are effective at capturing complex, non-linear behaviour but often lack transparency and generalizability, and their performance heavily depends on the quality and quantity of training data.\\
In contrast, the grey-box model strikes a balance between accuracy, interpretability, and practicality \cite{bacher2011identifying}. By combining simplified physical equations (e.g., \ac{RC} thermal models) with data-driven calibration, it allows us to effectively model building thermodynamics using readily available time-series data and minimal building metadata. This approach retains critical physical dynamics, provides explainable results, and supports adaptability for new scenarios. Grey-box modelling is also particularly suited to integration with \ac{MPC}, enabling real-time optimisation by leveraging dynamic predictions and constraints \cite{vsiroky2011experimental,privara2013building}.

We collect historical time-series data from the building's \ac{BMS} and incorporate weather data to fit the parameters of the \ac{RC} model to the historically observed data. Once fitted, the digital twin model can be initialised with a recently measured temperature, and used together with external forecasts in order to determine optimal heating and cooling inputs to maintain the desired internal temperature of the building zone for the next time horizon. The digital twin model is validated by simulating the existing legacy control of the building and comparing to the measured historic data.


The digital twin is then integrated with an optimisation model which works via \ac{MPC} approach. The optimisation model’s objective is to minimise energy consumption, carbon emission, energy cost, and the difference between the internal temperature and the target temperature of the building. Constraints are added such that the optimisation follows the thermodynamic rules of the building as set by the digital twin model, and so that it controls the \ac{HVAC} within the bounds of its given specifications. The inputs to the optimisation are the initial state of the building and \ac{HVAC}, and temporal energy prices and carbon intensity of the primary energy used to heat and cool the building. Optimisation weights, which allow the user to configure the balance between the objectives, are also an input.


The optimisation problem minimises the total cost, energy consumption, and carbon emissions while maintaining occupant comfort by keeping the internal temperature ($T_i$) close to the target set-point ($T_{\text{target}}$).
\[
\min \sum_{t=0}^{T-1} 
    w_{\text{cost}} \cdot C_{\text{energy}}(t) \cdot E_{\text{HVAC}}(t) 
    + w_{\text{carbon}} \cdot C_{\text{carbon}}(t) \cdot E_{\text{HVAC}}(t) 
    + w_{\text{comfort}} \cdot (T_i(t) - T_{\text{target}}(t)) 
\]
where $E_{\text{HVAC}}(t)$ is the primary energy consumption of \ac{HVAC} (heating and cooling) at time $t$, $C_{\text{energy}}(t)$ is the cost of energy at time $t$, $C_{\text{carbon}}(t)$ is the carbon dioxide emission factor at time $t$. $w_{\text{cost}},\, w_{\text{carbon}},\, w_{\text{comfort}}$ include a combined weight and scaling factor for cost, carbon dioxide emissions, and comfort deviation respectively.



To ensure that the predicted internal temperature follows the thermodynamic equations derived from the building's properties, we add thermodynamic constraints. The internal temperature ($T_i$) evolves according to the building's thermodynamics, which are derived from the discretised \ac{RC} thermal model \ac{ODE}. A simplified equation is given below, although the exact set of thermodynamic constraints and model order will depend on the building and the parameters determined as part of the model fitting pipeline.

\[
T_i(t+1)=T_i(t)+\frac{\Delta t}{C} \Big( \frac{T_e(t)-T_i(t)}{R}+Q_{\text{HVAC}}(t)+p\,I_s(t) \Big)
\]

where
$T_e$ denotes external temperature, $R$ denotes the thermal resistance of building zone, $C$ denotes the thermal capacitance of building zone,  $Q_{\text{HVAC}}$ denotes the heating or cooling power from the \ac{HVAC} system,  $I_s$ the solar irradiance,  $p$ is a parameter determining the factor of the solar irradiance incident on the zone and  $\Delta t$ denotes the timestep of the simulation.


The model respects the heating and cooling capacities of the \ac{HVAC} system, preventing decisions that exceed the system’s operational thresholds.
The \ac{HVAC} system operates within defined power limits:
\[
0\leq Q_{\text{HVAC}}(t)\leq Q_{\text{max}}
\]

Additional constraints can also be enabled to ensure the model respects the heating or cooling response of the real system, for example if weather compensation is used.


Soft constraints are added to penalise the model from deviating from the target set-point temperature (which is set by the building operator) during occupied hours, and to bias the optimisation to only allow \ac{HVAC} running during pre-set times if there are operational limitations.

Once the digital twin parameters are estimated and integrated with the \ac{MPC}-based optimisation model, the workflow proceeds in a rolling time horizon. At each step, the model predicts the building’s internal temperature for a future horizon (e.g. 24 hours) based on weather forecasts, energy prices, and \ac{HVAC} settings. The \ac{MPC} then determines the optimal \ac{HVAC} settings to maintain the desired temperature while minimising costs and carbon emissions.\\

\begin{figure}[h!]
    \centering
    \includegraphics[width=0.9\linewidth]{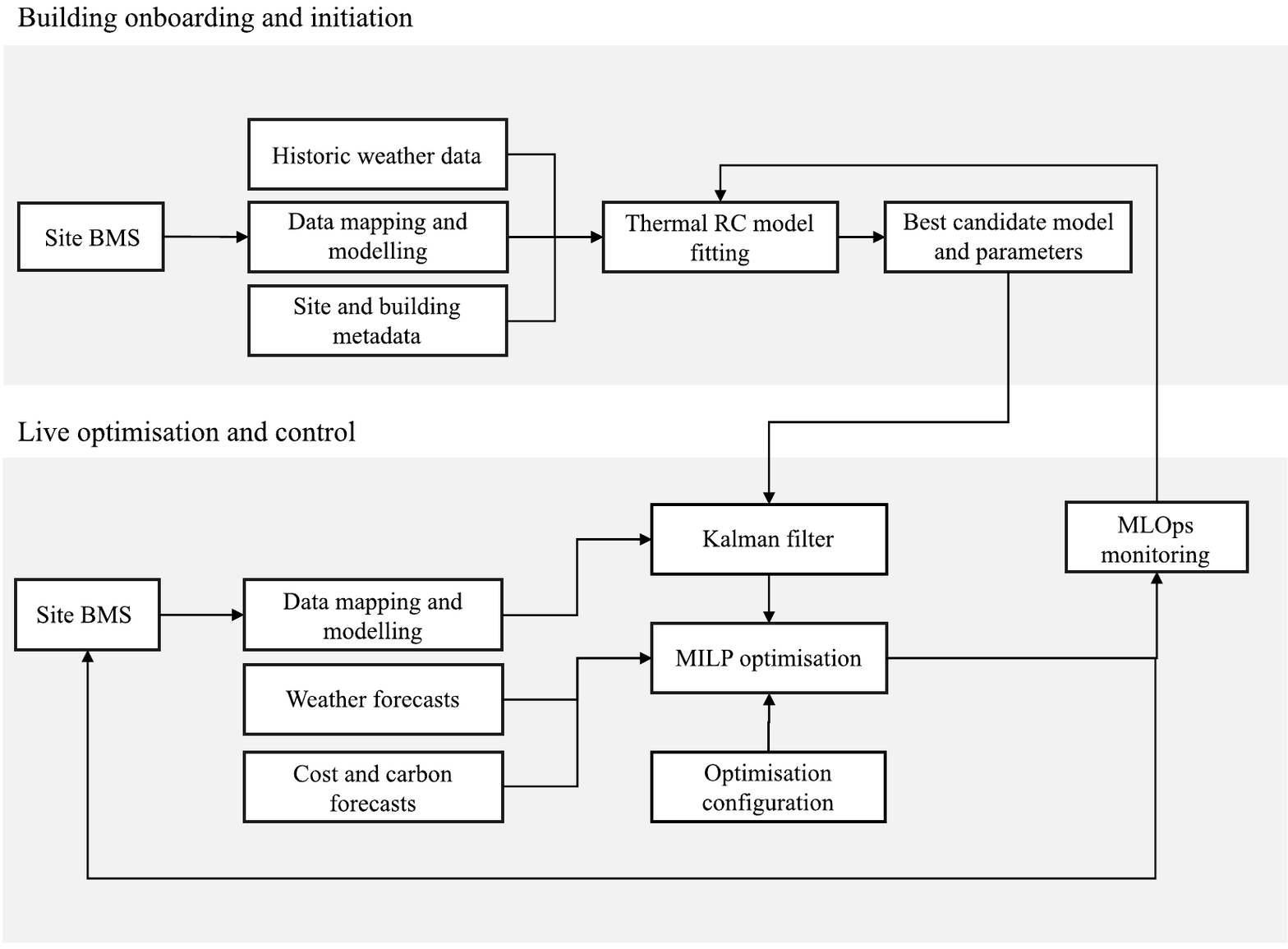}
    \caption{High-level schematic of the optimisation process}
    \label{figure_BEOPT1}
\end{figure}

A high-level schematic of optimisation process is shown in Figure~\ref{figure_BEOPT1}. One of the key strengths of the \ac{MPC} approach is its ability to adjust control actions in real-time based on new data. The external temperature, solar irradiance, and energy price forecasts can change frequently, affecting the building's thermodynamics and energy requirements. By updating the model at each time step with real-time data, the \ac{MPC} ensures that the optimisation remains relevant and effective despite dynamic conditions. Additionally, sensor readings from inside the building (for internal temperature) provide feedback to correct any deviations between predicted and actual states.

To further enhance the model's adaptability and handle uncertainties, a Kalman filter is employed within the system. The Kalman filter provides real-time estimation of unknown states, such as the latent temperature states of the \ac{RC} model or \ac{HVAC} system efficiency, by integrating sensor data with predictions from the model. This not only improves the accuracy of state estimation but also enables the system to adapt dynamically to drifting building behaviour, ensuring robust and reliable performance under varying conditions.

The complete solution is deployed on Databricks, with end-to-end \ac{ETL}, \ac{MLOps}, and monitoring to support model lifecycle management and continuous improvement. \ac{RC}-model parameters and optimisation inputs and results are tracked over time, allowing model performance and drift to be monitored. Retraining pipelines enable ``challenger" models to be promoted if significant model drift is detected, indicative of a physical or operational change in the building. \ac{ETL} pipelines are implemented big-data tools such as Apache Spark, data streaming, and medallion architecture - ensuring scalability and performance.

\subsection{Tests, Analysis and Discussion}

To evaluate the proposed building energy optimisation algorithms, a real-world test case was established using a commercial office building. The selected zone comprised an open-plan office space served by a central \ac{AHU}, which delivered both heating and cooling via gas-fired boilers and electric chillers. Hardware was deployed to site to enable a data pipeline to collect meter and sensor readings from over 200 individual zone and asset devices at 30-minute intervals. Using this data, predictive models were trained and optimisation routines developed. The optimisation objective aimed to maintain similar levels of occupant comfort during occupied hours while achieving cost and energy savings compared to the legacy control strategy.

\begin{figure}[h!]
    \centering
    \includegraphics[width=1.0\linewidth]{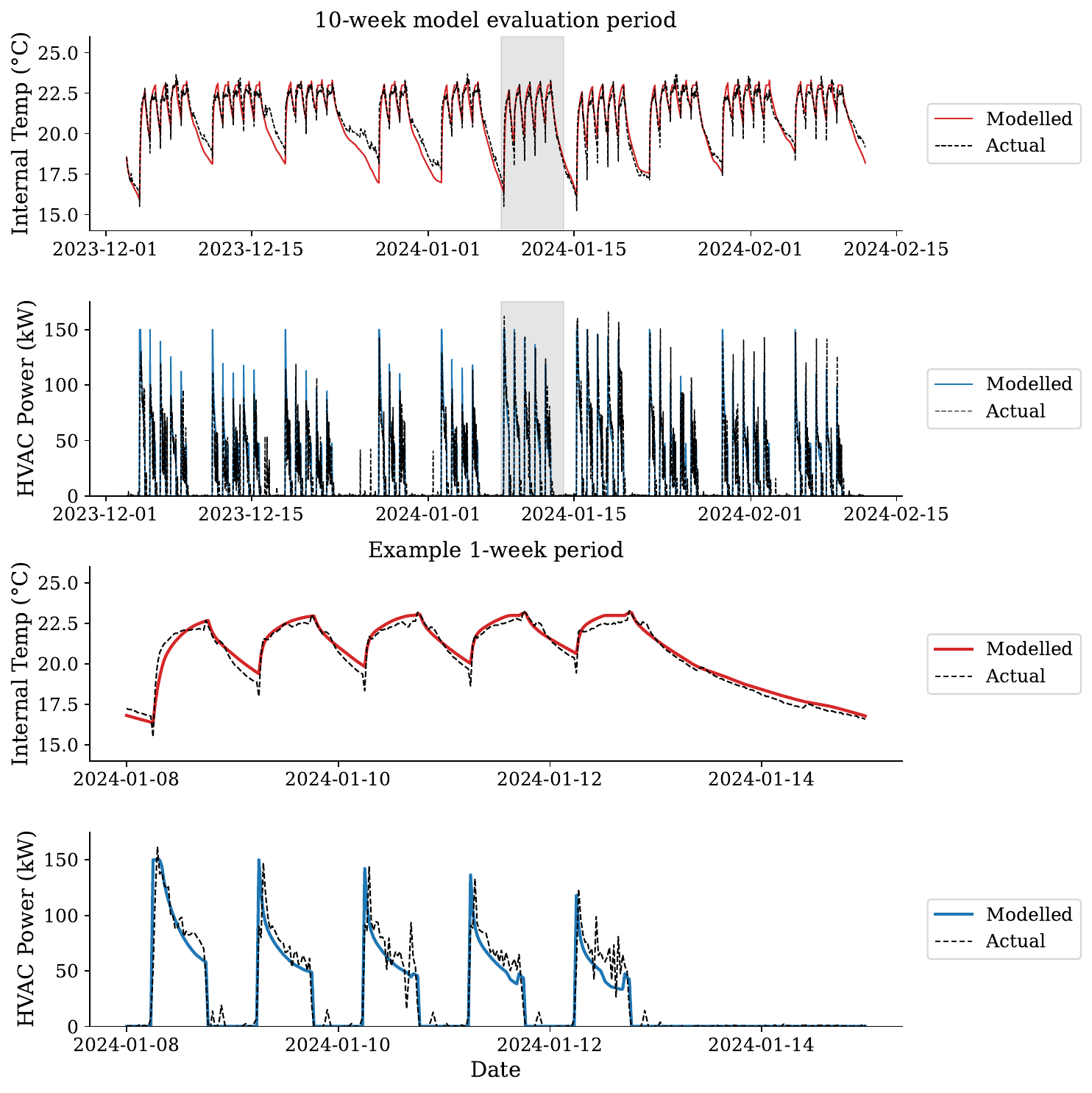}
    \caption{Model validation - simulation of the baseline strategy using the fitted model/digital twin vs. metered data}
    \label{figure_BEOPT2}
\end{figure}

Prior to implementing and deploying the optimisation, the model is validated using historic data. Figure~\ref{figure_BEOPT2} shows a baseline model validation - the simulated vs. actual internal temperature and heating power over a 10 week period. The baseline model validation for our test case achieved an R² value of 0.88 and 0.44°C \ac{MAE} between simulated and real temperatures. The simulated heating energy usage and cost was within 3.5\% of the metered quantity over the test period.

\begin{figure}[h!]
    \centering
    \includegraphics[width=1.0\linewidth]{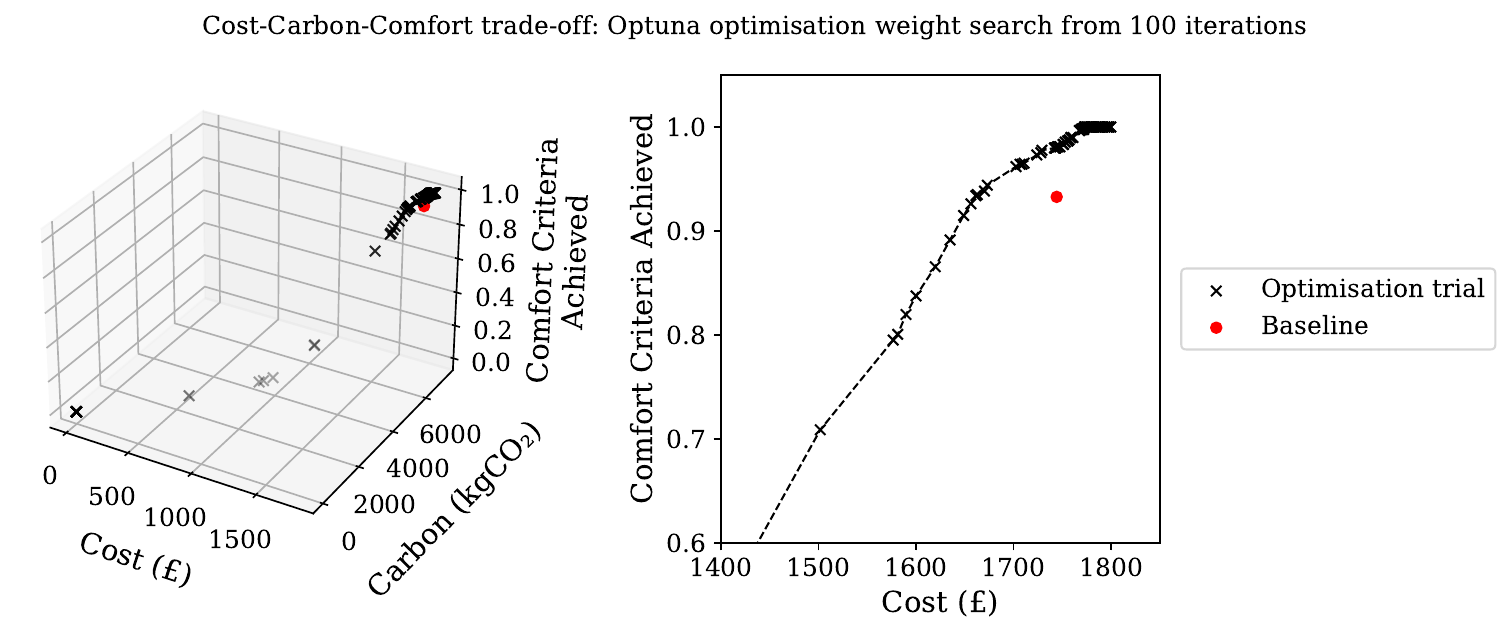}
    \caption{Pareto-front plots following hyper-parameter search on optimisation weights}
    \label{figure_BEOPT3}
\end{figure}

Following the model validation, a hyper-parameter search using the open-source \emph{Optuna} framework was conducted. Figure~\ref{figure_BEOPT3} shows the Pareto-front plot from running a hyper-parameter tuning experiment on the model. This enables the cost, carbon, and set-point deviation interaction of the solution space to be explored for different model weights. Comfort criteria refers to the proportion of occupied hours where the internal temperature of the space is within 1°C of the target set-point. This enables user interaction and model biasing. For this trial building and time period, a higher level of thermal comfort could be achieved for a similar energy cost by making time and temperature-based adjustments to the building set-point. Alternatively, a small energy cost saving (typically in range 3-8\% for this already well-optimised test building) could be made while maintaining a similar level of set-point compliance.

In addition to our real-world test case, proof-of-concept experiments were also carried out using open-source data from different types of buildings and \ac{HVAC} configurations. In these experiments, cost savings of up to 25\% were achieved, particularly in buildings more sensitive to solar gain and outside temperatures, or where the existing \ac{HVAC} is poorly controlled. The largest savings were observed where control could be optimised around dynamic time-of-use tariffs, for example by pre-heating or pre-cooling buildings during periods of lower energy costs.

This work demonstrates a robust and adaptive framework for real-time building energy optimisation, with a particular focus on managing uncertainties, integrating renewables, and ensuring practical deployment and monitoring. The use of Kalman filtering enables dynamic updating of building model parameters and latent states, allowing the system to adapt to long-term changes and seasonal variations. While \ac{MPC} offers strong control performance, future improvements could involve the use of stochastic \ac{MPC} to account for uncertainties in inputs such as weather forecasts and energy prices, enhancing system robustness. The integration of modern \ac{MLOps} and \ac{ETL} pipelines with proven techniques such as \ac{MPC} and Kalman filtering ensures the proposed framework can be scaled and deployed.

The optimisation approach also shows strong potential when integrated with on-site renewable energy and storage systems and time-of-use tariffs, allowing \ac{HVAC} operations to be aligned with solar generation and battery usage to reduce costs and grid dependence. Additional signals, such as demand-side response events (e.g. the Demand Flexibility Service in the UK) can also be integrated, enabling automated turn-down of \ac{HVAC} assets and an additional revenue source for building operators.

However, real-world deployment still poses several challenges. Integrating with diverse on-site systems requires reliable data pipelines, often involving complex and bespoke integration work. Data mapping is complicated by inconsistent or opaque metadata, necessitating standardisation efforts. Additionally, the availability and resolution of meter data can limit the granularity of optimisation, particularly in older buildings where metering infrastructure is limited. Finally, while grey-box \ac{RC} models are computationally efficient and physically interpretable, their linear nature may constrain their ability to fully capture complex thermal dynamics of some buildings.

Despite these challenges, the proposed system provides a scalable and practical foundation for advanced building energy optimisation, with clear pathways for enhancement through improved data infrastructure and ontology, richer modelling techniques, and tighter integration with emerging smart energy technologies.

\section{CHP-based Heat Network Optimisation}\label{section:CHOP}


Heat networks are now considered to be vital to the future of the UK energy landscape and for creating a pathway for the decarbonisation of heat at a large scale, which remains a challenge to the UK's net-zero targets. The UK Government has a target to increase the level of heat demand met by heat networks to 20\% by 2050. Particularly in high-density urban areas, they may provide the lowest cost and carbon solution to meeting heat demands for commercial and domestic premises.

While future heat networks will primarily use low-carbon generation sources such as heat pumps and waste heat, the majority of operational heat networks currently use gas-fired generation. Combined heat and power engines provide heat to their district networks, and electricity to both local private networks and for export.

The \ac{CHP} optimisation project aimed at advancing how \ac{CHP} systems are managed and operated. While \ac{CHP} can offer efficiency through co-generation, its full potential is often underutilised due to complex operational decisions - when to run the \ac{CHP} unit, how much electricity to export versus use on-site, and how to manage heat production and storage. \ac{CHP} and heat network optimisation addressed these challenges through an integrated optimisation framework, enabling operators to:

\begin{itemize}
    \item \textbf{Reduce operating costs:} Dynamically adjust operations based on fluctuating fuel and electricity prices, leveraging low-cost energy windows and minimising expensive peak imports.
    \item \textbf{Lower carbon emissions:} Ensuring existing gas-fired \ac{CHP} engines run only when cost and carbon-efficient, otherwise meeting demand with other heat sources.
    \item \textbf{Enhance reliability and data visibility:} Connecting existing plant and assets to cloud-based platforms enables closer to real-time data monitoring and fault and anomaly detection.
\end{itemize}

At its core, heat network optimisation implements white-box digital twin models of heat network sites and assets together with \ac{MILP} to optimise a range of variables - \ac{CHP} unit output, boiler usage, chiller settings, and energy storage operations. The resulting solution identifies an optimal strategy that balances cost, carbon, and operational constraints, translating directly into financial and environmental benefits and providing a price and carbon-aware supervising layer in addition to the local asset control. The heat network optimisation framework is designed to adapt to:

\begin{itemize}
    \item \textbf{Multiple energy sources and vectors:} \ac{CHP} engines, boilers, and chillers can be incorporated in order to meet site electricity, heat, hot water, and chilled water production.
    \item \textbf{Storage integration:} Thermal and electrical storage systems to smooth out supply-demand mismatches and enable decoupling of generation and demand.
    \item \textbf{Market and tariff interactions:} Imports and exports to the grid, as well as private wire agreements, ensuring profit maximisation and ensuring maximum export during periods of high wholesale prices.
    \item \textbf{Practical asset constraints:} Maintenance schedules, maximum daily restarts, ramp-up limits, and other real-world operational considerations.
\end{itemize}

\subsection{Methodology and Implementation}

The developed solution utilises white-box models of heat network assets such as \ac{CHP} engines, boilers, and chillers and incorporates these models into a \ac{MILP} framework which includes both physical and non-physical costs and constraints. Plant configuration and operational details are first collected in collaboration with asset managers and domain experts. These details are then translated into a set of mathematical constraints and an objective function, allowing optimisation by a computational solver to meet the needs of the business/site operator. A final stage involves transforming model outputs into implementable rules and schedules that can be read by the existing site hardware and control systems, enabling retrofitting to legacy assets.

The model comprises:

\begin{itemize}
    \item \textbf{Binary Variables:} Represent discrete actions (e.g., starting or stopping the \ac{CHP} units).
    \item \textbf{Schedule Variables:} Indicate operational status (on/off) of the plant.
    \item \textbf{Real Variables:} Capture continuous decisions, such as \ac{CHP} output percentages, grid import/export quantities, and levels of thermal storage.
\end{itemize}

A typical business use-case of the optimisation framework is to minimise the total operational cost of running the heat network, i.e. to minimise:

\[
\min \sum_{t=0}^{T-1} C_{\text{Real}}(t) + C_{\text{Artificial}}(t)
\]

The objective combines real costs (fuel, maintenance, and carbon) with artificial costs - penalties encouraging smoother operations without enforcing hard constraints.  \(C_{\text{Real}}\) includes production costs (i.e. those associated with generating electricity and heat) and maintenance costs - typically a function of the asset running hours.
\begin{align*}
\text{C}_{\text{Real}}(t) = {} &
    \text{G}_{\text{total}}(t) \cdot \text{P}_{\text{gas}}(t) \\
& + \text{E}_{\text{import}}(t) \cdot \text{P}_{\text{elec,import}}(t) \\
& - \text{E}_{\text{export}}(t) \cdot \text{P}_{\text{elec,export}}(t) \\
& + \text{C}_{\text{maintenance}}(t)
\end{align*}
where \(\text{G}_{\text{total}}(t)\) is the total gas consumed by the \ac{CHP} and boiler at time \(t\), 
\(\text{E}_{\text{import}}(t)\) is the electricity imported from the grid, and \(\text{E}_{\text{export}}(t)\) is the electricity exported to the grid. The corresponding price terms - \(\text{P}_{\text{gas}}(t) \), \(\text{P}_{\text{elec,import}}(t)\), and \( \text{P}_{\text{elec,export}}(t) \) - represent the unit prices of gas, electricity import, and electricity export respectively, all at time \( t \). \(\text{C}_{\text{maintenance}}\) is simplified as a fixed value at each timestep \(t\) representing the cost for running the site and its operational overheads.

Artificial costs (\(C_{\text{Artificial}}\)) are those that steer the optimisation toward operational best practices and smoother operation - like avoiding unnecessary boiler usage, limiting frequent restarts, or reducing simultaneous grid import and export.

\[
\text{C}_{\text{Artificial}}(t) = \text{C}_{\text{import, pref}}(t) + \text{C}_{\text{CHP, pref}}(t) + \text{C}_{\text{boiler, pref}}(t) + \text{C}_{\text{restart}}(t)
\]
Here, \( \text{C}_{\text{import, pref}} \) represents the penalty associated with importing electricity rather than generating it locally, aiding in smoother optimisation results and preventing import/export symmetry or oscillation. \( \text{C}_{\text{CHP, pref}} \) and \( \text{C}_{\text{boiler, pref}} \) enable the optimisation to be configured to prefer usage of certain CHPs (where there is more than one CHP unit) or boilers, and controls their interaction with thermal storage. \( \text{Cost}_{\text{restart}} \) enables penalisation of frequent plant restarts, which introduce additional wear and fatigue on the assets.

In addition to costs and penalties, constraints are added which enforce the model to respect energy conservation and physical asset limitations.  These include constraints enforcing that demand (both electricity and heat) is met at all timesteps:
\[
\text{E}_{\text{demand}}(t) = \text{E}_{\text{CHP}}(t) + \text{E}_{\text{import}}(t) - \text{E}_{\text{export}}(t)
\]
\[
\text{Q}_{\text{demand}}(t) = \text{Q}_{\text{CHP}}(t) + \text{Q}_{\text{Boilers}}(t) + \text{Q}_{\text{storage}}(t)
\]
Where \( \text{E}_{\text{demand}}(t) \) is the total electrical demand at time \( t \), \( \text{E}_{\text{CHP}}(t) \) is the electricity generated by the \ac{CHP} at time \( t \), \( \text{E}_{\text{import}}(t) \) is the electricity imported from the grid at time \( t \), and \( \text{E}_{\text{export}}(t) \) is the electricity exported to the grid or private wire at time \( t \). \( \text{Q}_{\text{demand}}(t) \) represents the total heat demand at time \( t \), \( \text{Q}_{\text{CHP}}(t) \) is the heat generated by the \ac{CHP}, \( \text{Q}_{\text{Boilers}}(t) \) is the heat generated by the boilers, and \( \text{Q}_{\text{storage}}(t) \) is the net heat discharged from thermal storage, all at time \( t \).

A further constraint of
\[
\text{E}_{\text{export}}(t) \le \text{E}_{\text{CHP, max}}(t) 
\]
ensures that the site cannot export more electricity than can be generated by the \ac{CHP} engines (for example, by simultaneous import and exporting electricity in an arbitrage manner). \( \text{R}(t) \) is a binary variable indicating whether a restart occurs at time \( t \), \( \mathcal{T}_{\text{period}} \) is the set of all timesteps within a given period (typically a day), and \( \text{R}_{\text{max}} \) is the maximum number of allowed restarts within that period. This ensures that the optimised solution restricts daily restarts or total operating hours, reflecting maintenance contracts or good practice based on the increased fatigue induced by frequent asset cycles.

\[
\sum_{t \in \mathcal{T}_{\text{period}}} \text{R}(t) \leq \text{R}_{\text{max}}
\]

Additional constraints on thermal storage ensure the model does not violate energy conservation laws or the capacity of the thermal stores.

\[
\text{SOE}_{\text{thermal}}(t+1) = \text{SOE}_{\text{thermal}}(t) + \text{Q}_{\text{storage}}(t)
\] subject to \(\text{SOE}_{\text{thermal, min}} \le \text{SOE}_{\text{thermal}}(t) \le \text{SOE}_{\text{thermal, max}}\) where $\text{SOE}_{\text{thermal}}$ refers to the state-of-energy of the thermal store in kWh and the subscripts $_{min}$ and $_{max}$ refer to minimum and maximum permitted level of the the thermal storage. 

\subsection{Tests, Analysis and Discussion}

The implementation as currently designed allows the optimal dispatch of an entire week’s \ac{CHP} schedule in a single, integrated process. This begins with the input of data, either forecasted or historical, covering a full seven-day period, including information on energy demand, market prices, and weather conditions. A single optimisation run is then performed using this data, allowing the solver to determine the most cost-effective sequence of operational decisions such as on/off states, output levels, and storage usage. The high level operational process and interaction between functions is shown below in Figure~\ref{figure_CHP1}.

\begin{figure}[h!]
    \centering
    \includegraphics[width=0.8\linewidth]{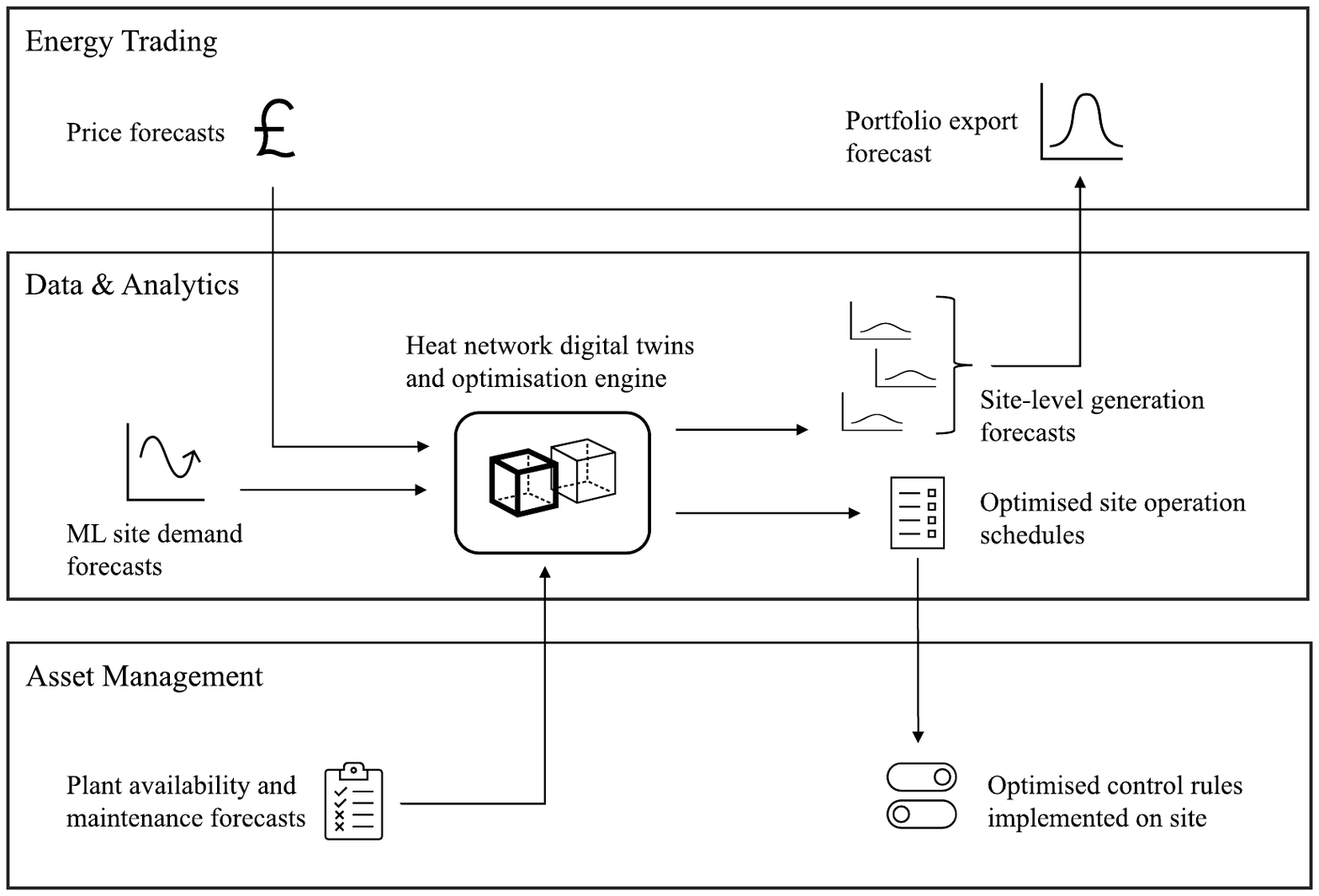}
    \caption{Heat network optimisation operational process}
    \label{figure_CHP1}
\end{figure}

\begin{figure}[h!]
    \centering
    \includegraphics[width=0.8\linewidth]{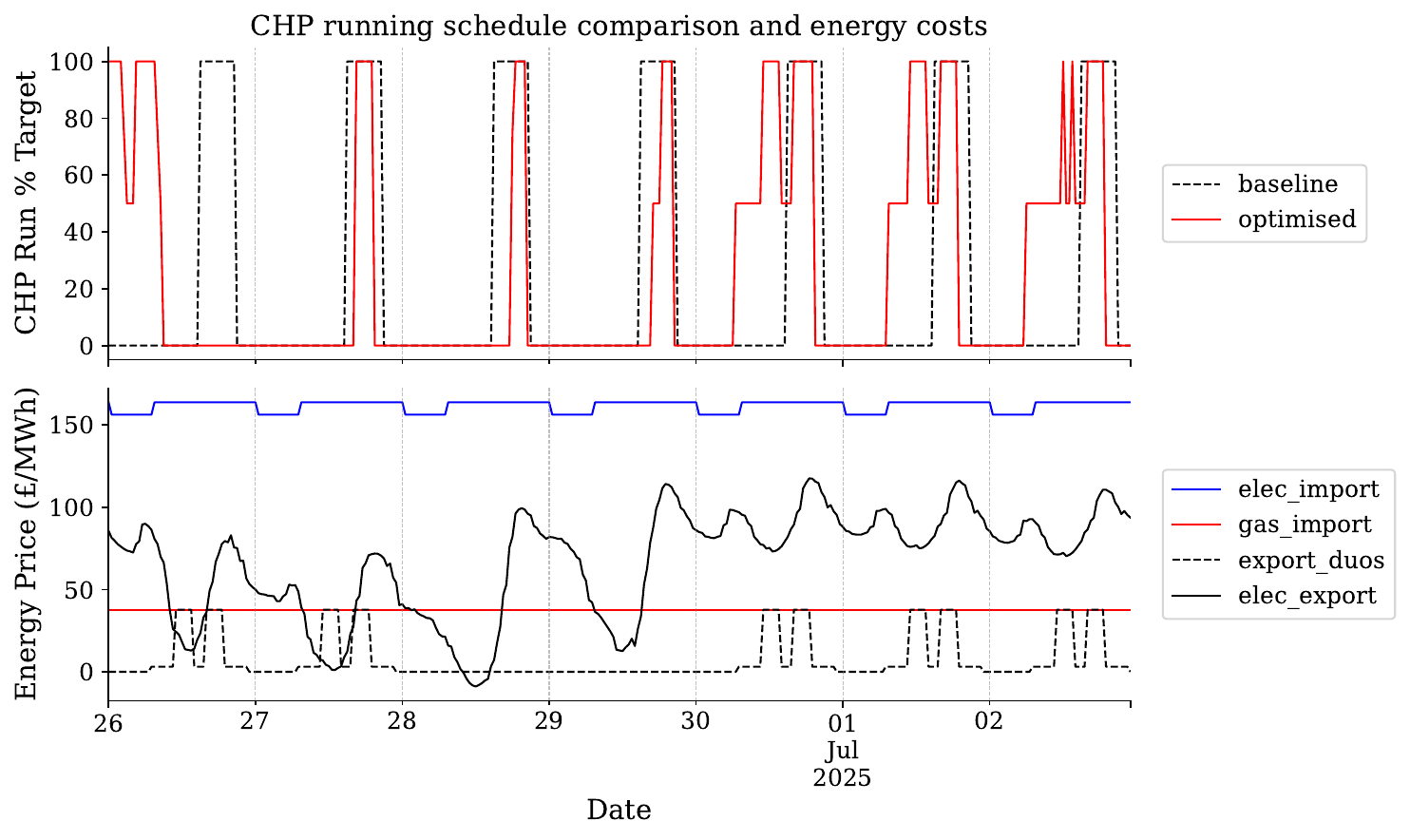}
    \caption{Optimised and baseline schedule comparison and week-ahead energy prices}
    \label{figure_CHP2}
\end{figure}

The resulting schedule offers a comprehensive guide for plant operators, enabling effective daily and hourly management with only minor manual adjustments necessary if real-world conditions diverge from projections. This weekly optimisation approach ensures that all decisions across the seven-day horizon are optimised for cost and operational constraints, making it highly valuable for strategic planning. An example optimised vs. baseline comparison is shown in Figure~\ref{figure_CHP2} for one of the test sites. While the baseline schedule aims to broadly capture the evening peak export prices, the optimised schedule is significantly more responsive to within-day price variation. This involves sometimes not running the \ac{CHP} if export prices are not favourable, and instead meeting heat demand using high-efficiency boilers, or running a reduced output and ramping up to meet price peaks or use-of-system incentives for generators.

From a financial and environmental standpoint, optimisation delivers several key benefits. Cost efficiency is achieved by minimising fuel consumption and aligning energy production with periods of lower market prices, potentially reducing operational costs by a significant percentage. This is particularly advantageous in the UK’s volatile energy market, where real-time responsiveness is crucial. Carbon emissions reduction is another major benefit; improved efficiency means lower greenhouse gas emissions, helping organizations meet sustainability goals in line with the UK’s net-zero targets, while also minimising exposure to carbon-related regulatory costs. Grid resilience and flexibility are also enhanced, as \ac{CHP} plants optimised through the process can better support both public and private energy networks by managing peak demands and reducing grid dependency, or reducing export during periods of grid congestion. 

Table~\ref{Table:CHOP1} and Table~\ref{Table:CHOP2} show the estimated running costs and the optimisation benefits on 3 trial sites where the model is operational. Estimated costs are the net of gas import cost, electricity import cost, maintenance cost, and export revenue (grid and private wire). Maintenance costs assume an hourly run cost for the \ac{CHP} based on asset team input and contracts. Each heat network site comprises a combination of residential and commercial heat and electricity demand, and are located in different regions of the UK. The cost analysis is based on three key cost categories: baseline, actual, and optimised. The baseline refers to the estimated costs of operating the site using the original or legacy schedules, typically fixed time windows and rules-based approaches such as \emph{run at 100\% between 7am to 12am}, as provided by the asset manager. These estimates are derived from the digital twin model that simulates system behaviour under those traditional schedules. The actual costs represent the real, calculated expenditure based on metered operational data from the site, reflecting how the system was actually run during the period. The optimised costs represent the theoretical minimum costs that could have been achieved if schedules and outputs were fully optimised in response to real-time factors such as heat demand, export prices, and other system constraints.

\begin{table}[htbp!]
\centering
\begin{threeparttable}
\caption{Estimated Running Costs }\label{Table:CHOP1}
\begin{tabular}{l l  r r }
\hline
 &  &  \multicolumn{2}{c}{Estimated Running   Cost Reduction \% :} \\

Site & Period &    Baseline $\rightarrow$ Actual & Baseline $\rightarrow$ Optimised  \\
\hline
Site 1 & 15 Apr 23 – 31 Mar 24 &  21\% & 22\% \\
Site 2 & 1 Dec 23 – 31 Mar 24 &  10\% & 19\%  \\
Site 3 & 1 Apr 23 – 31 Mar 24 &  15\% & 39\% \\
\bottomrule
\end{tabular}
\end{threeparttable}
\end{table}

\begin{table}[htbp!]
\centering
\begin{threeparttable}
\caption{ Optimisation Benefits}\label{Table:CHOP2}
\begin{tabular}{l l   r l l}
\hline
 &  &   \% Value Realised : & Site  &  Model \\

Site & Period &     Baseline $\rightarrow$ Optimised &  Complexity & Confidence \\
\hline
Site 1 & 15 Apr 23 – 31 Mar 24 &   96\% & L & H \\
Site 2 & 1 Dec 23 – 31 Mar 24 &   53\% & L & H \\
Site 3 & 1 Apr 23 – 31 Mar 24 &  38\% & H & L \\
\bottomrule
\end{tabular}
\end{threeparttable}
\end{table}

The analysis also examines two types of savings. The baseline to actual savings reflect the realised cost reductions achieved by moving away from fixed legacy schedules to the optimised schedules in use during the trial period. This gives an indication of the value already being captured through more dynamic operation and responsiveness to market prices and fluctuating demands. Meanwhile, the baseline to optimised savings represent the maximum potential cost savings that could be realised under a fully automated, ``perfect world'' optimisation scenario - i.e. using automated dispatch and assuming perfect forecast accuracy. This highlights the remaining opportunity for improvement in the process, e.g. by moving from a week-ahead to day-ahead optimisation horizon. Together, these comparisons provide a clear view of both progress to date and remaining opportunities for cost reduction.

In addition to improved automation, future improvements could draw on techniques introduced in the previous sections, improving demand forecasting by integrating external features such as weather and other exogenous factors as well as tighter control through \ac{MPC}. Similarly, integrating more granular data, such as  \ac{IoT}/\ac{LoRaWAN} meters on key customer branches would decrease demand forecasting volatility, particularly if combined with short-horizon/nowcasting models that capture building inertia and occupant patterns. Looking forward, integrating the ability to perform multi-objective optimisation against conflicting goals, e.g. cost, carbon footprint, and service-level \acp{KPI} would further increase the flexibility of this system.

\section{Energy Management System Optimisation in a System-of-Systems Context}\label{section:ems}

This section introduces a white-box approach to \ac{EMS} optimisation, combining \ac{MILP} with transparent \acp{DT} of solar \ac{PV}, wind, and \ac{BESS}. The methodology is demonstrated through a study conducted at a school in London, with an emphasis on its broader applicability to microgrid and \ac{SoS} architectures. The solution features a rolling-horizon optimisation framework that accommodates time-varying electricity tariffs, battery cycling limitations, and optional peak-shaving constraints. Central to the design are two optimisation components: a Digital Twin Baseline for rule-based reference control with the option of a Self-Consumption optimiser that prioritises local energy use, and a Cost-Based optimiser that targets economic efficiency. The model also supports comparative scenario analysis across different tariff structures, for example, flat and peak/off-peak pricing, enabling tailored strategies based on local market conditions.

This asset-by-asset digital-twin and MILP structure can simply be cloned or linked together to manage multiple sites as one system‐of‐systems. Each site’s generation, storage and load models can feed into the same rolling-window optimiser, which allows for aggregate forecasts and constraints across systems (e.g. buildings or microgrids) without changing the core solution. In practice, this allows for scaling from a single building to a cluster of buildings (or any other set of locations) by adding more data streams.

A key focus of the work is on accurate and physically interpretable asset modelling, as well as the integration of real sensor data through robust data ingestion pipelines. The interplay between power-based constraints (e.g. kW peak limits) and energy-based decision variables (e.g. kWh storage dispatch) is examined in depth, highlighting the operational trade-offs in constrained environments. Preliminary results indicate cost savings of 8 to 12\% compared to rule-based strategies (with particularly strong savings achieved when combined with time-of-use tariffs) along with additional benefits in carbon intensity reduction, demonstrating the dual financial and environmental value of advanced \ac{EMS} strategies. While the pilot does not disclose the site’s identity beyond noting its urban school setting, the approach is readily transferable to other contexts, including those with wind generation or additional distributed energy resources. Overall, the proposed framework provides a scalable, transparent, and data-driven \ac{EMS} solution that balances physical realism with computational efficiency.

\subsection{Methodology and Implementation}

Although the method supports solar, wind, and load management, our pilot site primarily features:
\begin{itemize}
    \item \textbf{Solar \ac{PV} array} of moderate capacity,
    \item \textbf{\ac{BESS}} in the 100\,kWh class,
    \item Local building load in the range of tens of kW peak,
    \item \textbf{Time-varying tariffs} for grid import and a fixed export rate for export.
\end{itemize}

The broader project includes two phases:
\begin{enumerate}
    \item \textbf{Phase 1 \ac{PoC}}: Develop the core optimisation engine and demonstrate its feasibility using historical data for a single site. 
    \item \textbf{Phase 2 \ac{MVP}}: Generalise the model to any site configuration (solar, wind, battery, flexible loads), integrate live data streams, and deploy a stable service for real-time or day-ahead scheduling.
\end{enumerate}
We focus primarily on Phase~1 but highlight generalisation aspects (e.g.\ multiple tariff scenarios) that pave the way for Phase~2.  For the Solar Digital Twin, we simulate half-hourly (or user-selected granularity) generation, $E_{\text{PV}}(t)$, with a function $f_{\text{PV}}$:
\[
  E_{\text{PV}}(t) = f_{\text{PV}}\bigl(I(t),\, T_{\text{ambient}}(t)\bigr),
\]
calibrated to panel \ac{DC} ratings, efficiency curves, and simple temperature de-rating. Forecasted global horizontal irradiance $I(t)$ is combined with ambient temperature, $T_{\text{ambient}}(t)$, and other local adjustments (tilt, orientation, cloud coverage) to estimate energy output. An optional night-time force-zero logic ensures that no negative or spurious generation is reported between dusk and dawn.

For the Battery Digital Twin, energy stored in the battery unit is measured as the State of Charge,  $\text{SOC}(t)$,  the percentage of the nominal energy capacity, $E_{\text{cap}}$, utilised at a given time. The battery is treated as a generator, so the energy discharged over time, $E_{\text{Discharge}}(t)$ ,  is treated as positive and the energy stored in the battery, $E_{\text{Charge}}(t)$ , negative. Accounting for losses in the storage process, charging and discharging efficiencies, $\eta_{\text{c}}$ and $\eta_{\text{d}}$, are also applied.
In each discrete interval $\Delta t$ (e.g.\ 30\,minutes):
\begin{equation}
  \text{SOC}(t+1) 
    = \text{SOC}(t)\;+\;
      \bigl(\eta_{\text{c}}\,E_{\text{Charge}}(t)\;-\;\frac{E_{\text{Discharge}}(t)}{\eta_{\text{d}}}\bigr)
      \,\times\,\frac{100}{E_{\text{cap}}}\,,
  \label{eq:soc_update}
\end{equation}
subject to $\text{SOC}_{\min}\,\le\,\text{SOC}(t)\,\le\,\text{SOC}_{\max}$. Further, per-interval power constraints:
\[
  E_{\text{Charge}}(t) \le \text{P}_{\max}, 
  \quad 
  E_{\text{Discharge}}(t) \le \text{P}_{\max}.
\]
A cycle count limit can be enforced via the sum of $(E_{\text{Charge}}(t) + E_{\text{Discharge}}(t)) / (2\,E_{\text{cap}})$ over the horizon.
\\\\
We considered two main \ac{EMS} strategies:
\begin{enumerate}
  \item \textbf{Baseline (Digital Twin with Self-Consumption option):}  
    A direct, rule-based reference combining:
    \begin{itemize}
      \item \emph{Digital Twin:}  
        Charge the battery whenever $E_{\text{PV}}(t) > E_{\text{Load}}(t)$ (up to SOC and power limits), and discharge whenever $E_{\text{PV}}(t) < E_{\text{Load}}(t)$, using the continuous SOC update of equation~\eqref{eq:soc_update} and the energy balance.
      \item \emph{Self-Consumption:}  
        Enforce via MILP
        \[
          E_{\text{Charge}}(t)\le\max\{E_{\text{PV}}(t)-E_{\text{Load}}(t),0\},\quad
          E_{\text{Discharge}}(t)\le\max\{E_{\text{Load}}(t)-E_{\text{PV}}(t),0\},
        \]
        thus ensuring the battery only charges from local renewables and only discharges to serve load. An optional \texttt{peak\_shaving\_mode} adds
        \[
          E_{\text{Import}}(t)\le E_{\text{threshold}}
        \]
        per interval. No grid-to-battery charging or price arbitrage is allowed.
    \end{itemize}

  \item \textbf{Cost Optimisation:}  
    A fully flexible MILP that minimises
    \[
      \sum_{t=0}^{T-1}
      \Bigl(
        \text{P}_{\text{import}}(t)\,E_{\text{Import}}(t)
        - \text{P}_{\text{export}}\,E_{\text{Export}}(t)
        + w_{\text{carbon}}\,I_{\text{GridCO2}}(t)\,E_{\text{Import}}(t)
      \Bigr),
    \]
    subject to the SOC evolution (equation~\eqref{eq:soc_update}), power limits, and
    \[
      E_{\text{Load}}(t)
      = E_{\text{PV}}(t)
        + E_{\text{Import}}(t)
        - E_{\text{Export}}(t)
        + E_{\text{Discharge}}(t)
        - E_{\text{Charge}}(t).
    \]
    Charging from the grid is allowed for price arbitrage, and an optional cycle-count constraint can protect battery longevity.
\end{enumerate}

\subsection{Tests, Analysis and Discussion}

The study explores \ac{EMS} optimisation under various tariff structures beyond a single import rate, including flat rates, peak/off-peak, three-tier time-of-use, and weekday-only peak tariffs. Scenario-specific cost modelling was achieved by dynamically adjusting time-varying import prices, with sensitivity analyses conducted on base price levels, peak multipliers, and export rates. These analyses revealed how tariff differentials significantly influence the economic value of battery storage, offering insights particularly relevant to microgrids and energy communities aiming to maximise returns. A robust data ingestion pipeline ensures reliable forecasts and system alignment, critical for real-world deployment.

\begin{figure}[h!]
    \centering
    \includegraphics[width=1.0\linewidth]{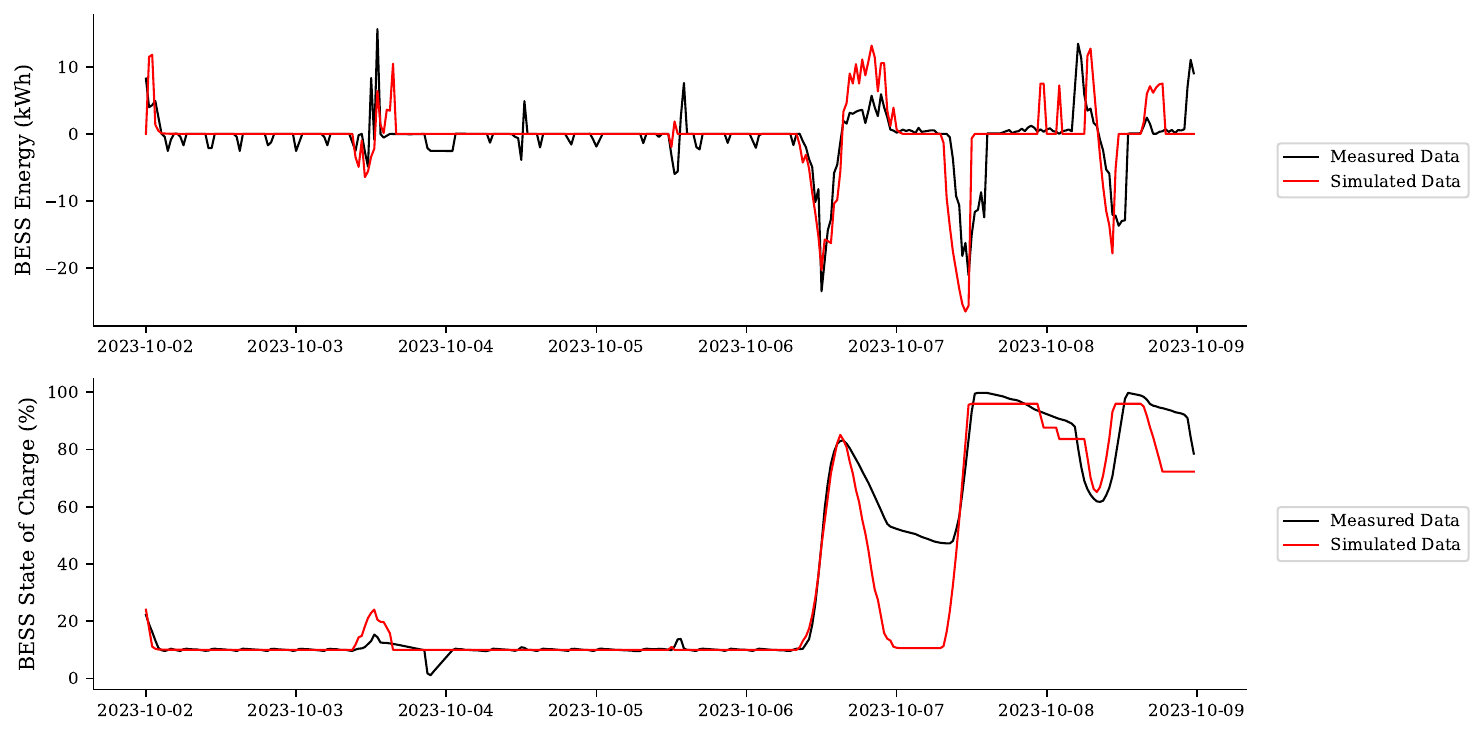}
    \caption{Battery DT Evaluation}
    \label{figure_Batt1}
\end{figure}

Figure~\ref{figure_Batt1} presents the evaluation of the battery digital twin under ideal solar conditions, where the solar input corresponds to actual measured data. The performance assessment is based on averaged results across four randomly selected weeks in both summer and winter. For the Day-Ahead Modelling, the digital twin achieved a high level of accuracy, with an R² score of 0.97, a \ac{NMAE} of 0.04, and a symmetric Mean Absolute Percentage Error \ac{sMAPE} of 2\%. The associated uncertainty in cost impact was less than 1\%. In comparison, the Week-Ahead Modelling showed slightly lower predictive performance, with an R² score of 0.95, \ac{NMAE} of 0.1, \ac{sMAPE} of 6\%, and a corresponding cost impact uncertainty of approximately 3\%. These results highlight the strong predictive capabilities of the battery digital twin, particularly in shorter forecasting horizons.

\begin{figure}[h!]
    \centering
    \includegraphics[width=1.0\linewidth]{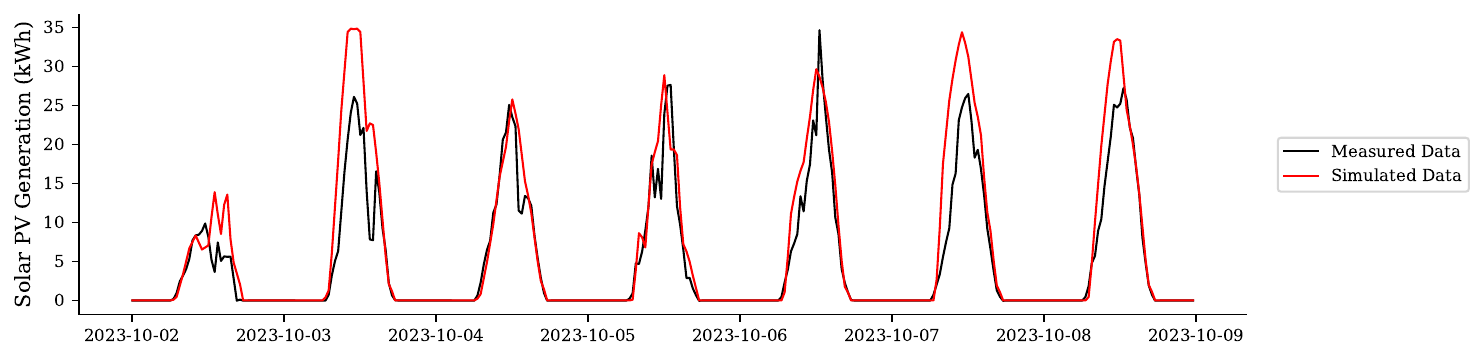}
    \caption{Solar DT Evaluation}
    \label{figure_Batt2}
\end{figure}

Figure~\ref{figure_Batt2} illustrates the evaluation of the solar digital twin (\ac{DT}) under ideal battery energy storage system (\ac{BESS}) conditions, using actual \ac{BESS} measurement data but low-quality solar irradiance inputs. The analysis is based on average results across four randomly selected weeks in both summer and winter. Under these conditions, the Day-Ahead Modelling achieved an R² score of 0.95, a \ac{NMAE} of 0.09, a \ac{sMAPE} of 4\%, and an associated cost impact uncertainty of approximately 2\%. For the Week-Ahead Modelling, the predictive accuracy slightly decreased, with an R² score of 0.93, \ac{NMAE} of 0.19, \ac{sMAPE} of 10\%, and a cost impact uncertainty of around 6\%. These results indicate that while the solar \ac{DT} remains reasonably robust even with lower-quality irradiance data, forecasting accuracy and cost reliability are notably more sensitive over longer horizons.

\begin{figure}[h!]
    \centering
    \includegraphics[width=1.0\linewidth]{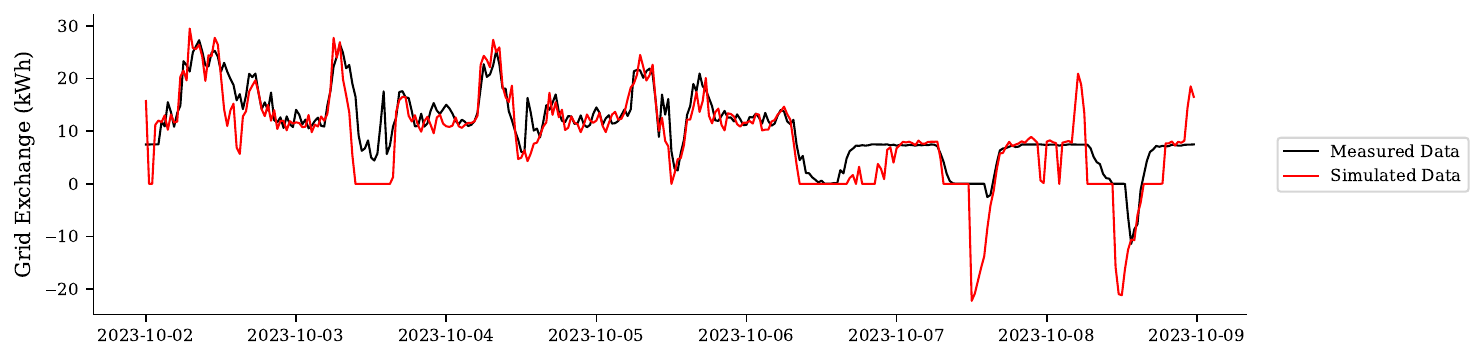}
    \caption{EMS DT Evaluation}
    \label{figure_Batt3}
\end{figure}

Finally, Figure~\ref{figure_Batt3} presents the evaluation of the Energy Management System (\ac{EMS}) digital twin. The results are based on average values computed across four randomly selected weeks in both summer and winter. For the Day-Ahead Modelling, the \ac{EMS} digital twin achieved an R² score of 0.93, a \ac{NMAE} of 0.12, a \ac{sMAPE} of 5\%, and an estimated cost impact uncertainty of approximately 5\%. In comparison, the Week-Ahead Modelling showed a moderate reduction in accuracy, with an R² score of 0.90, \ac{NMAE} of 0.21, \ac{sMAPE} of 6\%, and a corresponding cost impact uncertainty of around 10\%. These results suggest that while the \ac{EMS} digital twin performs well under both forecasting horizons, the reliability of predictions and cost estimates declines slightly with longer-term forecasts.

Our comparisons between baseline (with and without self-consumption), and fully cost-optimised strategies showed that while self-consumption simplifies operation, cost-optimised control—using arbitrage—can lead to greater savings, particularly under dynamic tariffs. At the pilot school site in London, cost reductions ranged from 3–5\% under simple tariffs to 8–12\% with aggressive time-of-use structures. Small carbon savings (0.5–1\%) were also observed when incorporating carbon intensity signals. Peak-shaving introduced challenges due to mismatches between power-based constraints (kW) and the model’s energy-based (kWh) scheduling, highlighting the need for finer control granularity or integration with local controllers. Practical challenges such as data quality, forecast errors, and scalability were also noted; future expansions may require real-time re-optimisation, sub-interval scheduling, and hierarchical or distributed control to ensure performance and responsiveness at scale.

\clearpage

\section{Conclusion and Future work}\label{section:conclusion}

This paper has shown that next-generation forecasting and optimisation methods can deliver sizeable efficiency, carbon-reduction and resilience gains across four tightly linked arenas: high-resolution, weather-enhanced demand forecasting; grey-box + MPC optimisation of individual buildings; CHP-centred heat-network dispatch; and whole-site Energy Management System optimisation within a broader System-of-Systems (SoS) architecture. Field deployments on UK sites confirmed that ensemble learning markedly sharpened forecast accuracy, AI/ML-driven control trimmed consumption without sacrificing comfort, and coordinated EMS schedules simultaneously lowered cost and $\mbox{CO}_2$ by responding in real time to price and grid-carbon signals. 

This was achieved through two main technical approaches. First, the work built end-to-end data pipelines that blend meteorological, sensor and market information, ensuring a single, reliable source for both short-horizon control and long-horizon planning. Second, the framework made effective use of planning and optimisation strategies through Model Predictive Control and Mixed Integer Linear Programming.   These techniques effectively closed the gap between asset-level optimisation and site-wide or even portfolio-wide value. The same workflow proved transferable from a single commercial building to a district-heating scheme and a school micro-grid, demonstrating that a modular approach can scale with minimal re-engineering effort.

Future research should first explore the feasibility of employing current state-of-the-art AI/ML black-box paradigms for (spatio)-temporal data, such as Temporal Fusion Transformers and graph neural networks, as well as emerging grey-box paradigms from physics-informed machine learning.  On the optimisation and planning side,  scenario-based and distributionally robust MPC frameworks should be explored to ensure robustness of schedules  against inevitable uncertainties in weather, prices and equipment availability, ensuring that optimisation remains feasible.

Deployment architecture also needs to edge outward. Pushing selected inference and control routines to on-site or roadside edge devices will enable sub-second actuation, respect data-sovereignty requirements and open the door to federated learning schemes that share model improvements without moving raw data. At the same time, the SoS layer should widen its scope to embrace electric-vehicle fleets, hydrogen production and storage, industrial demand-response assets and long-duration thermal stores, making sector coupling an operational reality.

Sustainability metrics must move beyond operational kilowatt-hours. Integrating lifecycle and Scope-3 carbon accounting into optimisation objectives will keep decisions aligned with net-zero goals even when embodied emissions dominate.   More broadly, progress will be faster if the community converges on open, modular ecosystems, such as the ones detailed in this paper. Contributing to standards such as IEC 61850 extensions and OpenADR, publishing deployment blueprints and offering explainable-AI dashboards can all cut integration time and build trust. Embedding human-in-the-loop design and studying social-behaviour feedbacks, regulatory incentives and community-energy models will ensure that technically elegant solutions translate into real-world adoption.

Taken together, these directions chart a path from isolated optimisation pilots to self-learning, secure and socially inclusive energy ecosystems—a critical stride toward meeting the United Kingdom’s 2035 carbon-reduction milestones and supporting the global drive for net-zero.

\section*{Acronyms}\begin{acronym}[htp]

\acro{AHU}{Air Handling Unit}
\acro{ARIMA}{Autoregressive Integrated Moving Average}
\acro{BESS}{Battery Energy Storage Systems}
\acro{BMS}{Building Management System}
\acro{CHP}{Combined Heat and Power}
\acro{DC}{Direct Current}
\acro{DER}{Distributed Energy Resource}
\acro{DL}{Deep Learning}
\acro{DT}{Digital Twin}
\acro{EMA}{Exponential Moving Average}
\acro{EMS}{Energy Management System}
\acro{ETL}{Extract, Transform, Load}
\acro{HVAC}{Heating, Ventilation, and Air-Conditioning}
\acro{IoT}{Internet of Things}
\acro{KF}{Kalman Filter}
\acro{KPI}{Key Performance Indicator}
\acro{LightGBM}{Light Gradient-Boosting Machine}
\acro{LoRaWAN}{Long Range Wide Area Network}
\acro{LP}{Linear Programming}
\acro{LSTM}{Long Short Term Memory}
\acro{MAE}{Mean Absolute Error}
\acro{MAPE}{Mean Absolute Percentage Error}
\acro{MILP}{Mixed-Integer Linear Programming}
\acro{ML}{Machine Learning}
\acro{MLOps}{Machine learning operations}
\acro{MPAN}{Meter Point Administration Number}
\acro{MPC}{Model Predictive Control}
\acro{MVP}{Minimum Viable Product}
\acro{NMAE}{Normalised Mean Absolute Error}
\acro{ODE}{Ordinary Differential Equation}
\acro{PoC}{Proof of Concept}
\acro{PV}{Photovoltaic}
\acro{RC}{Resistance-Capacitance}
\acro{RL}{Reinforcement Learning}
\acro{RMSE}{Root Mean Square Error}
\acro{RNN}{Recurrent Neural Network}
\acro{SoS}{Systems-of-Systems}
\acro{sMAPE}{symmetric Mean Absolute Percentage Error}
\acro{XGBoost}{eXtreme Gradient Boosting}

\end{acronym}

\newpage

\section*{List of Variables}

\begin{align*}
T_i(t) &\quad \text{Internal temperature at time } t \\
  T_{\text{target}}(t) &\quad \text{Target internal temperature at time } t \\
  T_e(t) &\quad \text{External temperature} \\
  E_{\text{HVAC}}(t) &\quad \text{HVAC energy consumption at time } t \\
  I_s(t) &\quad \text{Solar irradiance} \\
  Q_{\text{HVAC}}(t) &\quad \text{HVAC heating/cooling power} \\
  R, C &\quad \text{Thermal resistance and capacitance} \\
  C_{\text{energy}}(t), C_{\text{carbon}}(t) &\quad \text{Cost and carbon factors} \\
  w_{\text{cost}}, w_{\text{carbon}}, w_{\text{comfort}} &\quad \text{Weights for multi-objective optimisation}\\
G_{\text{total}}(t) &\quad \text{Total gas consumed} \\
  E_{\text{import}}(t), E_{\text{export}}(t) &\quad \text{Electricity imported/exported} \\
  Q_{\text{CHP}}(t), Q_{\text{Boilers}}(t), Q_{\text{thermal}}(t) &\quad \text{Heat produced by respective sources} \\
  P_{\text{gas}}(t), P_{\text{elec,import}}(t), P_{\text{elec,export}}(t) &\quad \text{Price terms} \\
  R(t) &\quad \text{Restart indicator (binary variable)} \\
  \text{SOE}_{\text{thermal}}(t) &\quad \text{State of energy of thermal storage} \\
  \text{Charge}_{\text{thermal}}(t), \text{Discharge}_{\text{thermal}}(t) &\quad \text{Thermal storage charge/discharge}\\
  SOC(t) &\quad \text{Battery state-of-charge} \\
  \text{Charge}(t), \text{Discharge}(t) &\quad \text{Battery charging/discharging rates} \\
  \eta_c, \eta_d &\quad \text{Battery charging/discharging efficiency} \\
  E_{\text{cap}} &\quad \text{Battery energy capacity} \\
  G_{\text{PV}}(t) & \quad \text{PV generation function} \\
  \text{CO2Intensity}(t) &\quad \text{Carbon intensity of grid power} \\
  \hat{y}^{(m)}_{t+h} & \quad \text{Forecast from model } m \text{ at time } t+h \\
  w_m & \quad \text{Weight assigned to model } m \\
  \hat{y}^{\text{(ens)}}_{t+h} & \quad \text{Ensemble forecast}
\end{align*}

\appendix
\include{appendix}

\bibliographystyle{alpha}
\bibliography{sample}

\end{document}